\documentstyle[12pt]{article}

\begin{document}
\thispagestyle{empty}
\begin{center}
{\Large {\bf Scissors Modes and Spin Excitations in Light Nuclei
including $\Delta N$=$2$ excitations:}}

\vspace{0.2in}

{\Large {\bf Behaviour of $^8Be$ and $^{10}Be$}}

\vspace{0.3in}

M. S. Fayache, S. Shelly Sharma* and L. Zamick

\begin{small}
{\it Department of Physics and Astronomy, Rutgers University,
	Piscataway, NJ 08855}
\end{small}

\vspace{0.2in}

\begin{small}
{*\it Departamento de F\'{i}sica, Universidade Estadual de Londrina,
Londrina, Parana, 86051-970, Brazil}
\end{small}

\end{center}

\begin{abstract}
Shell model calculations are performed for magnetic dipole excitations
in $^8\mbox{Be}$ and $^{10}\mbox{Be}$ in which all valence
configurations plus $2\hbar\omega$ excitations are allowed (large
space). We study both the orbital and
spin excitations. The results are compared with the `valence space
only' calculations (small space). The cumulative energy weighted sums
are calculated and compared for the $J=0^+$ $T$=$0$ to $J=1^+$ $T$=$1$
excitations in $^8\mbox{Be}$ and for $J=0^+$ $T$=$1$ to both $J=1^+$
$T$=$1$ and $J$=$1^+$ $T$=$2$ excitations in $^{10}\mbox{Be}$. We
find for the $J=0^+$ $T$=$1$ to $J=1^+$ $T$=$1$ isovector {\underline {spin}}
transitions in $^{10}\mbox{Be}$ that the summed strength in the
{\underline {large}} space is less than in the {\underline {small}}
space.  We find that the high energy energy-weighted isovector orbital
strength is smaller than the low energy strength for transitions in
which the isospin is changed, but for $J=0^+$ $T$=$1$ to $J=1^+$
$T$=$1$ in $^{10}\mbox{Be}$ the high energy strength is larger. We
find that the low lying orbital strength in $^{10}\mbox{Be}$ is
anomalously small, when an attempt is made to correlate it with the
$B(E2)$ strength to the lowest $2^+$ states. On the other hand a sum
rule of Zheng and
Zamick which concerns the total $B(E2)$ strength is reasonably
satisfied in both $^8\mbox{Be}$ and $^{10}\mbox{Be}$. The Wigner
supermultiplet scheme is a useful guide in analyzing shell model
results. In $^{10}Be$ and with a $Q \cdot Q$ interaction the $T=1$ and
$T=2$ scissors modes are degenerate, with the latter carrying
$\frac{5}{3}$ of the $T=1$ strength.
\end{abstract}

\pagebreak

\section{The Experimental Situation}

\hspace{.25 in}
{}From our perspective, much experimental information is lacking in the
nuclei $^8\mbox{Be}$ and $^{10}\mbox{Be}$. For example, no $J=1^+$
states have been identified in $^{10}\mbox{Be}$. The $B(E2)$ from the
$2_1^+$ state of $^8\mbox{Be}$ to the $J=0^+$ ground state is not
known -this is understandable because of the large decay width to two
alpha particles.

The following states and their properties are of interest to us:

{\bf {(a) $^8\mbox{Be}$}}

The $J=2_1^+$ state has an excitation energy of $3.04$ $MeV$. The
$J=4_1^+$ state is at $11.4$ $MeV$. This is consistent with an $J(J+1)$
spectrum of a rotational band, but it should be recalled that any
spin-independent interaction gives an $J(J+1)$ spectrum in the $p$
shell. The $J=1_1^+$ $T=1$ state, which we discussed extensively in a
previous publication \cite{fay} is at $17.64$ $MeV$ and the $J=1_1^+$
$T=0$ state is at $18.15$ $MeV$.

The $B(M1)$ from the $17.64$ $MeV$ state to the ground state has a
strength of $0.15$ $W.u.$ or $B(M1)$$\put(2,8.){\vector(0,-9){9}}$ $=$
$0.27{\mu_N}^2$. The $B(M1)$ of this state to the $2_1^+$ state is
$0.12$ $W.u.$ or $B(M1)$$\put(2,8.){\vector(0,-9){9}}$ $=$
$0.21{\mu_N}^2$ \cite{ajz}. Of course $B(M1)\uparrow$=$3B(M1)\downarrow$

{\bf {(b) $^{10}\mbox{Be}$}}

The $J=2_1^+$ state is at $3.368$ $MeV$ and the $J=2_2^+$ state at
$5.960$ $MeV$. We recall that with a spin independent interaction the
$2_1^+$ and $2_2^+$ would be degenerate. The experimental spectrum
looks more vibrational. However, the values of
$B(E2)$$\put(2,0.){\vector(0,9){9}}$ from the $J=0^+$ ground state to
the $2_1^+$ state is very strong: $B(E2)$$\put(2,0.){\vector(0,9){9}}$
$=$ $52$ $e^2fm^4$. Raman et. al. deduce from this a deformation
parameter $\beta$ $=$ $1.13$ \cite{ram}. As mentioned above, there
are no $J=1^+$ states mentioned in the compilation of Ajzenberg-Selove
\cite{ajz}. Also the $J=4^+$ state has not been found.

\section{The Interactions}

\hspace{.25 in}
We have chosen two types of interactions to do the calculations. First
we use a short range `simplified realistic' ($x,y$) interaction
previously used by Zheng and Zamick \cite{ann}, and then we use a long-range
quadrupole-quadrupole interaction. By choosing these two extremes, we
make sure that the results we obtain are not too dependent on the
specifics of the model.

In more detail, the($x,y$) Hamiltonian  is:

\[H=\sum T_i + \sum_{i<j}V(ij)\]

\noindent where

\[V=V_c+xV_{so}+yV_t\]

\noindent with $c \equiv$central, $s.o. \equiv$spin-orbit, and $t
\equiv$tensor.

For ($x,y$)=(1,1) the matrix elements of this interaction are close to
those of realistic G matrices such as Bonn A. We can study the effects
the spin-orbit and tensor interactions by varying $x$ and $y$.

Note that we do not add any single-particle energies to the above
Hamiltonian. Rather, we let the single-particle energies be implicitly
generated by $H$. Hence, if we set $x$=0 i.e. turn off the
{\underline{two-body}} spin-orbit interaction, we will also be turning
off the one-body spin-orbit splitting coming from this interaction.

As a counterpoint, we repeat all the calculations with the $Q \cdot Q$
Hamiltonian

\[H_{Q}=\sum_i \frac{p_i^2}{2m} + \frac{1}{2}m\omega^2r_i^2
-\chi\sum_{i<j}Q \cdot Q\]

\noindent Note that we have added the term $\frac{1}{2}m\omega^2r^2$
which is not present for the ($x,y$) interaction. The reason for this
is that $Q \cdot Q$ cannot generate any single-particle potential energy
splitting whereas the ($x,y$) interaction can.

Whereas the ($x,y$) interaction like all realistic interactions is of
short range, the $Q \cdot Q$ interaction is long range. Yet, as we shall see
some of the results (but not all) are rather similar for the two
interactions. Since the best milieu for the existence of scissors mode
excitations (orbital magnetic dipole excitations) are strongly
deformed systems, one would expect the $Q \cdot Q$ Hamiltonian to yield
strong scissors modes. But is this also true for the realistic
interaction ? We will address this question. Another motivation for
introducing the $Q \cdot Q$ Hamiltonian is that it is easy to establish a
connection via energy weighted sum rule techniques between isovector
orbital $B(M1)$'s and isoscalar and isovector $B(E2)$'s.

We shall be performing the calculations, not only in the
$0\hbar\omega$ space (small space) but also in a space which allows
$2\hbar\omega$ admixtures (large space). For the $Q \cdot Q$ Hamiltonian in
the small space the energy matrix is proportional to $\chi$. Hence the
energy eigenvalues depend linearly on $\chi$, but the eigenfunctions
(and $B(M1)$'s and $B(E2)$'s) are independent of the interaction strength.
In a large space calculation there is one more parameter: the energy
splitting induced by $\frac{p^2}{2m} + \frac{1}{2}m\omega^2r^2$ i.e.
$2\hbar\omega$. Thus the wave function and the corresponding $B(M1)$'s
and $B(E2)$'s will also depend on $\chi$.

We have chosen values of $\chi$ appropriate for the large space
calculation. We also use these same values in the small space. One can
argue that in the small space one should use a renormalized value
$\chi'$ which is close to twice $\chi$. However, as mentioned above,
the wave function and hence $B(M1)$ and $B(E2)$ will not change, only
the energies. By choosing the same $\chi$ in the two spaces it is
easier to see what the differences in the two calculations are. The
values of $\chi$ are 0.5762 $\frac{MeV}{fm^4}$ for $^8Be$ and 0.3615
$\frac{MeV}{fm^4}$ for $^{10}Be$.

\section{The summed magnetic dipole strength}

\hspace{.25 in}
In Table I we give the summed magnetic dipole strength ($\sum_i~ B(M1:
0_1^+,~T=1 \rightarrow 1_i^+,~T=1)$ and $\sum_i~ B(M1:
0_1^+,~T=1 \rightarrow 1_i^+,~T=2)$ ) broken up into
isoscalar and isovector and spin and orbit and where we use the
($x,y$) interaction with $x=1$, $y=1$. We first discuss the
behaviour as a function of the size of the model space. Later we will
make a comparison of the behaviour in $^8\mbox{Be}$ and
$^{10}\mbox{Be}$. There are striking differences for the two nuclei.

Our small space calculation consists of all configurations of the form
$(0s)^4(0p)^4$ for $^8\mbox{Be}$ and $(0s)^4(0p)^6$ for
$^{10}\mbox{Be}$. The large space consists of those configurations
plus $2\hbar\omega$ excitations. Thus one can either excite two
particles to the next major shell or excite one particle through two
major shells. We also give results for the summed strength in the
{\it low-large space} -this is the low energy part of the large space
covering an energy range more or less equal to that of the small
space. It is easy to identify the low energy sector because there is a
fairly wide plateau in the summed strength which separates the low
energy rise from the high energy rise.

Usually the large space summed strength is somewhat larger than the
small space strength e.g. for the isovector orbital strength in
$^8\mbox{Be}$ the values shown in Table $I$ are $0.6701$ $\mu_N^2$ and
$0.7283$ $\mu_N^2$ respectively. But there is one glaring exception.
For the case of $J^{\pi}=0^+$ $T=1$ $\rightarrow$ $J^{\pi}=1^+$ $T=1$
transitions in $^{10}\mbox{Be}$, the summed isovector spin strength in
the large space is $2.08~10^{-2}$ $\mu_N^2$ but in the small space it
is {\underline {bigger}} $2.34~10^{-2}$ $\mu_N^2$. For the orbital
strength it is the other way around but for the physical case
($g_{l\pi}=1$, $g_{l\nu}=0$, $g_{s\pi}=5.586$, $g_{s\nu}=-3.826$) the
spin prevails and the summed strength in the {\underline {large space}}
$1.952$ $\mu_N^2$ is less than in the {\underline {small space}}
$2.09$ $\mu_N^2$.

Thus it is not always true that the net result of higher shell
admixtures is to rob strength from the low energy sector and move it
to higher energies. In some cases the total strength gets depleted.

We next compare the low energy sum in the large space with the small
space sum. In all cases the latter is larger than the low energy sum,
thus indicating that there is a quenching of the low energy part due
to higher shell admixtures. The hindrance factor $[(low~large)/small]$
is $0.88$ for the isovector orbital in $^8\mbox{Be}$, $0.73$ for the
isovector spin in $^8\mbox{Be}$, $0.77$ for the total $M1$ in
$^{10}\mbox{Be}$ etc.

Note that the total $M1$'s for $^{10}\mbox{Be}$ are somewhat larger
than for $^8\mbox{Be}$. However there is a dramatic drop in the
orbital strength in $^{10}\mbox{Be}$ relative to that in
$^8\mbox{Be}$. The large space summed orbital strength for
$^8\mbox{Be}$ is $0.73$ $\mu_N^2$ whereas for $^{10}\mbox{Be}$ (to
$J=1^+$ $T=1$ and $T=2$) the value is ($0.196~+~0.183$)=$0.38$
$\mu_N^2$. In the low energy sector the $^8\mbox{Be}$ value is $0.59$
$\mu_N^2$ whereas for $^{10}\mbox{Be}$ it is $0.10~+~0.13~=~0.23$
$\mu_N^2$, less than half the value for $^8\mbox{Be}$.

{}From the systematics of orbital transitions in heavy nuclei one
concludes that the proper milieu for isovector orbital transitions is
strongly deformed nuclei. Can one conclude that $^{10}\mbox{Be}$ is
not strongly deformed? The answer, by examining the tables of Raman
et. al. \cite{ram} is no! There is a strong $E2$ connecting the $0_1^+$
and $2_1^+$ states in $^{10}\mbox{Be}$. From this the authors conclude
that the deformation parameter $\beta$ is about $1.13$ -quite enormous.
Of course $^8\mbox{Be}$ might have an even stronger $E2$ transition
-there is no data on this in the Raman paper \cite{ram}, probably
because of the rapid decay of the $2_1^+$ state into two alpha
particles.

\section{The cumulative energy weighted strength for orbital
transitions in $^8\mbox{Be}$ and $^{10}\mbox{Be}$}

\hspace{.25 in}
\indent In this section we present results and figures for the cumulative
energy weighted sum of magnetic dipole strength.

We are motivated in so doing by various energy-weighted sum rules that
have been developed e.g. by Zheng and Zamick \cite{zz}, Heyde and de
Coster \cite{ibm}, Moya de Guerra and Zamick \cite{dz1}, Nojarov
\cite{no}, Hamamoto and Nazarewicz \cite{hz} and Fayache and Zamick
\cite{fay}. We will focus in particular on the orbital strength for
which the operator is ($\vec{L_{\pi}}-\vec{L_{\nu}}$)/2. In a previous
publication we presented results for the ($x,y$) interaction with
$x$=$1$, $y$=$1$ for $^8\mbox{Be}$ \cite{fay}. In this work the
quadrupole-quadrupole interaction results are compared with the
($x,y$) interaction results, and furthermore we extend the calculation
to $^{10}\mbox{Be}$. In the latter nucleus one does not have $N=Z$ and
this leads to big differences.

Whereas in $^8\mbox{Be}$ there is only one isospin channel for
isovector transitions $J=0_1^+$ $T=0$ $\rightarrow$ $J=1^+$ $T=1$, in
$^{10}\mbox{Be}$ there are two: $J=0_1^+$ $T=1$ $\rightarrow$ $J=1^+$
$T=1$ and $J=0_1^+$ $T=1$ $\rightarrow$ $J=1^+$ $T=2$. The low lying
$J=1^+$ $T=1$ states in $^{10}\mbox{Be}$ are expected to have much
smaller excitation energies than the $J=1^+$ $T=1$ states in
$^8\mbox{Be}$. This makes it easier to look for such states
experimentally.

In Table II we present the results for the summed energy weighted strengths
for the ($x,y$) interaction. As a crude orientation it should be noted
that simple models e.g. the Nilsson model used by de Guerra and Zamick
\cite{dz1} and a model by Nojarov \cite{no} would have the `large'
result be twice the `low large' result. On the other hand Hamamoto and
Nazarewicz \cite{hz} have argued that the `large' result should be
much more than twice the `low large' result. The actual ratios for the
($x,y$) and $Q \cdot Q$ interactions for this calculation (all
$0\hbar\omega$ configurations plus $2\hbar\omega$ excitations) are

\begin{small}
\noindent
\begin{center}
\begin{tabular}{cccc}
   &   & \multicolumn{2}{c}{RATIO}\\
    &      &  ($x,y$) & $Q \cdot Q$\\
$^8Be$ \hspace{0.2in} & $J=0^+$ $T=0$ $\rightarrow$ $J=1^+$ $T=1$ &
1.75 & 1.37 \\
$^{10}Be$ \hspace{0.2in}& $J=0^+$ $T=1$ $\rightarrow$ $J=1^+$ $T=2$ &
2.00 & 1.52\\
$^{10}Be$ \hspace{0.2in}& $J=0^+$ $T=1$ $\rightarrow$ $J=1^+$ $T=1$ &
3.22 & 3.68\\
$^{10}Be$ \hspace{0.2in}& (`$T=1$' + `$T=2$') & 2.52 & 2.33\\
\end{tabular}
\end{center}
\end{small}

\noindent For $^{10}Be$ we should actually compare the theoretical
models with the combined result (`$T=1$' + `$T=2$').

These results indicate that the simple models are not too bad as a
first orientation but there are fluctuations -sometimes the ratio is
less than two, sometimes greater. We will discuss these matters in
more detail in the context of the figures.

 In Figs $1$ and $2$ we show the cumulative energy
weighted {\underline {isovector orbital}} strength distributions in
$^8\mbox{Be}$ for the ($x,y$) interaction and for the Hamiltonian
$H_{Q}$ i.e. the quadrupole-quadrupole interaction. The results for the
two interactions are quite similar. The outstanding feature is that
there are two rises separated by a rather wide plateau. For the
($x,y$) interaction the first rise is to a plateau at about $12$
$\mu_N^2MeV$ followed by a second rise to about $20.8$ $\mu_N^2MeV$.
For the $Q \cdot Q$ interaction the first plateau is at $10.25$ $\mu_N^2MeV$
and the second at $14$ $\mu_N^2MeV$. A simple self-consistent Nilsson
model was shown to give the second plateau at twice the energy of the
first plateau \cite{zz} \cite{ibm}. That is to say the high energy
rise was equal to the low energy rise. In the more detailed
calculations performed here the high energy rise is less than the low
energy rise.

We next turn to $^{10}\mbox{Be}$. Here there are two channels:
$J=0^+$ $T=1$ $\rightarrow$ $J=1^+$ $T=1$ and $J=0^+$ $T=1$
$\rightarrow$ $J=1^+$ $T=2$. Let us discuss the  latter channel
first. The behaviour for $T=2$ in $^{10}\mbox{Be}$ is similar to that
for $T=1$ in $^8\mbox{Be}$. As shown in Figs $3$ and $4$ for the
($x,y$) and $Q \cdot Q$ interactions respectively, there are two rises
separated by a plateau and here the second rise is about twice the
first rise for the ($x,y$) interaction. For the $Q \cdot Q$
interaction (with $\chi=0.3615$) the low energy rise is to $1.7$
$\mu_N^2MeV$ and the next rise is to $2.6$ $\mu_N^2MeV$ -only $1.5$ to
one.

There is a big difference in the cumulative energy weighted
distributions, shown in Figs 5 and 6, for the $J=0^+$ $T=1$
$\rightarrow$ $J=1^+$ $T=1$ channel. For the ($x,y$)
interaction the first plateau (at about $2.5$ $\mu_N^2MeV$) is not
very flat, but the most outstanding feature in the curve is that the
high energy rise is much larger than the low energy rise. The energy
weighted sum reaches up to about $8$ $\mu_N^2MeV$. Thus the high
energy rise is over three time the low energy rise. For the $Q \cdot Q$
interaction, the first plateau is better defined -it is located at $1$
$\mu_N^2MeV$ and the cumulative sum extends to about $3.8$
$\mu_N^2MeV$.

\section{The Zheng-Zamick Sum Rule}

\hspace{.25 in}
Energy weighted sum rules for magnetic dipole transitions, be they
spin or orbital, are highly model dependent. An energy weighted sum
rule for {\it isovector orbital} magnetic dipole transitions for the
quadrupole-quadrupole interaction $Q \cdot Q$ was developed by Zheng and
Zamick \cite{zz}. This was motivated by the work of the Darmstadt
group \cite{zr} \cite{rich} showing a linear relationship between
summed orbital $B(M1)$ strength and the square of the deformation
parameter i.e. $\delta^2$.

\noindent The result was

\[\sum_n (E_n-E_0)B(M1)_o=\frac{9\chi}{16\pi}\left\{\sum_i[B(E2,0_1
\rightarrow 2_i)_{isoscalar} - B(E2,0_1 \rightarrow 2_i)_{isovector}]
\right\}~~~~(EWSR)\]

\noindent where $B(M1)_o$ is the value for the isovector orbital $M1$
operator ($g_{l\pi}=0.5$ $g_{l\nu}=-0.5$ $g_{s\pi}=0$ $g_{s\nu}=0$)
and the operator for the $E2$ transitions is $\sum_{protons}e_pr^2Y_2$
$+$ $\sum_{neutrons}e_nr^2Y_2$ with $e_p=1$, $e_n=1$ for the isoscalar
transition, and $e_p=1$, $e_n=-1$ for the isovector transition.

Let us now describe in detail how this sum rule works. The sum rule
should work for single-shell as well mixed-shell space.

We first consider the case of $^8\mbox{Be}$. We have the following
values in a large space calculation for the $H_{Q}$
interaction corresponding to orbital $M1$ excitations from the $J=0^+$
$T=0$ ground state to all $1^+$ $T=1$ states:

\begin{tabbing}
\= 1. Energy weighted isovector orbital $M1$ strength: \hspace{.7in}
\= {\underline {$14.040$}} \= $\mu_N^2MeV$\\

\> 2. The isoscalar summed strength $B(E2;1,1)$: \> $237.46$ \> $e^2fm^4$\\

\> 3. The isovector summed strength $B(E2;1,-1)$: \> $89.611$ \> $e^2fm^4$\\

\> 4. The right hand side ($\frac{9\chi}{16\pi}=0.1032$):
\> {\underline {$15.25$}} \> $\mu_N^2MeV$.\\
\end{tabbing}

We don't get exact agreement ($14.04$ $\mu_N^2MeV$ vs. $15.25$
$\mu_N^2MeV$) but it is reasonably close. One possible reason for the
disagreement is that spurious states have been removed and/or that
only $2\hbar\omega$ excitations to the $\Delta N=2$ shell are taken
into account \cite{ibm} \cite{zr}.

There have been other approaches, especially in the context of the
Interacting Boson Model \cite{ibm} which relate the energy weighted
orbital magnetic sum to the $B(E2)$ of the lowest $2^+$ state. As a
matter of curiosity we shall examine in our calculation what happens
if we take only the lowest $2^+$ state in the right hand side of the
sum rule ($EWSR$).

For the $2_1^+$ state in $^8\mbox{Be}$ we obtain (in our large space
calculation) $B(E2;1,1)=196.76$ and $B(E2;1,-1)=0$ (because the
$2_1^+$ state has $T=0$). The right hand side becomes $20.30$
$\mu_N^2MeV$. We get a {\underline{larger}} answer using the lowest
$2^+$ state than we do if we use all $2^+$ states in the
$0\hbar\omega$ and $2\hbar\omega$ space. The reason for this is that
when the lowest $2^+$ state is excluded, the isovector $B(E2)$ is
larger than the isoscalar $B(E2)$.

Can we also get agreement for the sum rule in the small space
$(0s)^4(0p)^4$ for $^8\mbox{Be}$? Now the numbers are:

\begin{tabbing}
\= 1. Energy weighted isovector orbital $M1$ strength: \hspace{1.in}
\= {\underline {$6.411$}} \= $\mu_N^2MeV$  \\

\> 2. The isoscalar summed strength $B(E2;1,1)$: \> $72.54$ \> $e^2fm^4$\\

\> 3. The isovector summed strength $B(E2;1,-1)$: \> $10.37$ \> $e^2fm^4$\\

\> 4. The right hand side ($\frac{9\chi}{16\pi}=0.1032$): \>
{\underline {$6.414$}} \> $\mu_N^2MeV$.\\
\end{tabbing}

We get perfect agreement.

We next consider $^{10}\mbox{Be}$. The relevant numbers for the large
space calculation are:

\begin{tabbing}
\= 1. Energy weighted $M1$ strength: \= $J=0^+,~T=1$ $\rightarrow$
$J=1^+,~T=1$ \hspace{0.5in}\= $ 3.811 $ \= $\mu_N^2MeV$ \\

\>  \> $J=0^+,~T=1$ $\rightarrow$ $J=1^+,~T=2$ \> $ 2.602 $ \> $\mu_N^2MeV$ \\

\>  Left Hand Side \> \> {\underline {$ 6.413 $}} \> $\mu_N^2MeV$ \\

\> 2. $\sum B(E2;1,1)$ \> $J=0^+,~T=1$ $\rightarrow$ $J=2^+,~T=1$ \>
$251.4$ \> $e^2fm^4$\\

\> 3.(a) $\sum B(E2;1,-1)$ \> $J=0^+,~T=1$ $\rightarrow$ $J=2^+,~T=1$
\> $94.02$ \> $e^2fm^4$\\

\> ~~(b) $\sum B(E2;1,-1)$ \> $J=0^+,~T=1$ $\rightarrow$ $J=2^+,~T=2$
\> $48.78$ \> $e^2fm^4$\\

\> ~~(c) $\sum B(E2;1,-1)$ \> Total \> $142.8$
\> $e^2fm^4$\\

\> 4. Right hand Side ($\frac{9\chi}{16\pi}=0.0647$): \>  \>
{\underline {$7.029$}} \> $\mu_N^2MeV$\\
\end{tabbing}

For $^{10}\mbox{Be}$ we are also curious to see what happens if
we use only the lowest $2^+$ state in the right hand side of the sum
rule. But we have to be careful! It turns out
that there is substantial $B(E2)$ strength to the two lowest $J=2^+$
states. This can be understood from the fact that with a $Q \cdot Q$
interaction in a small space calculation ($(0s)^4(0p)^6$) the two
lowest $2^+$ states are exactly degenerate. The states belong to the
$[f]=[42]$ representation. The $Q \cdot Q$ interaction fails to remove
the degeneracy of these states. Another way of stating this is that
the ($\lambda \mu$) values for both states are (22), and the allowed
values of the $K$ quantum number in the Nilsson scheme are $K=\mu$,
$\mu-2$, etc. Thus the $2^+$ states have $K=0$ and $K=2$.

When we go to the large space calculation with a $Q \cdot Q$ interaction,
limiting the excitations to $2\hbar\omega$, the degeneracy is removed
but the states are still fairly close together. The calculated values
are:

\begin{tabbing}
\= \hspace{2.in} \= $E2(1,1)$  \hspace{.5in}\= $E2(1,-1)$\\
\> $2^+_1$ \hspace{.5in} $E=2.08$ $MeV$ \> 64.94 \> 12.32\\
\> $2^+_2$ \hspace{.5in} $E=2.92$ $MeV$ \> 93.38 \> 10.11\\
\end{tabbing}

\noindent Thus, using the calculated values of $B(E2)$ for the lowest
two $2^+$ $T=1$ states in $^{10}\mbox{Be}$, we get for the right hand
side of the sum rule a value of $8.80$ $\mu_N^2MeV$. Again, as in the
case of $^8\mbox{Be}$, this is larger than the value $7.03$
$\mu_N^2MeV$ that is obtained by using all $2^+$ $T=1$ and all $2^+$
$T=2$ states.

The corresponding numbers in small space for $^{10}\mbox{Be}$ are:

\begin{tabbing}
\= 1. Energy weighted $M1$ strength: \= $J=0^+,~T=1$ $\rightarrow$
$J=1^+,T=1$ \hspace{0.5in}\= $ 0.7597 $ \= $\mu_N^2MeV$ \\

\>  \> $J=0^+,~T=1$ $\rightarrow$ $J=1^+,~T=2$ \> $1.266$ \> $\mu_N^2MeV$ \\

\>  Left Hand Side \> \> {\underline {$ 2.026 $}} \> $\mu_N^2MeV$ \\

\> 2. $\sum B(E2;1,1)$ \> $J=0^+,~T=1$ $\rightarrow$ $J=2^+,T=1$ \> $68.31$
\> $e^2fm^4$\\

\> 3.(a) $\sum B(E2;1,-1)$ \> $J=0^+,~T=1$ $\rightarrow$ $J=2^+,~T=1$
\> $33.80$ \> $e^2fm^4$\\

\> ~~(b) $\sum B(E2;1,-1)$ \> $J=0^+,T=1$ $\rightarrow$ $J=2^+,T=2$ \> $3.203$
\> $e^2fm^4$\\

\> ~~(c) $\sum B(E2;1,-1)$ \> Total \> $37.003$
\> $e^2fm^4$\\

\> 4. Right hand Side ($\frac{9\chi}{16\pi}=0.0647$): \>  \>
{\underline {$2.026$}} \> $\mu_N^2MeV$\\
\end{tabbing}

\section{A discussion of the calculated B(E2) values}

\hspace{.25 in}
Although the main thrust of this work is on $B(M1)$ values, we have
established a connection with $B(E2)$ for the orbital case. It is
therefore appropriate to discuss the calculated $B(E2)$ values
-comparing the behaviours in $^8\mbox{Be}$ and $^{10}\mbox{Be}$, and
comparing the different interactions that have been used (see tables
$III$ and $IV$.)

In making the comparison between $^8\mbox{Be}$ and $^{10}\mbox{Be}$ we
should lump together the $B(E2)$'s of the first two $2^+$ states in
$^{10}\mbox{Be}$ because with the interactions used here -especially
$Q \cdot Q$- these states are nearly degenerate. (However, experimentally the
states are well separated $E_{2_1^+}= 3.368$ $MeV$ and $E_{2_2^+}= 5.958$
$MeV$). When this is done we find that the $B(E2)$ values in the two
nuclei are comparable.

For the ($x,y$) interaction, the calculated (large space) value of
$B(E2)$ to the lowest two $2^+$ states in $^{10}\mbox{Be}$ is $22.90$
$e^2fm^4$, whereas it is $25.97$ $e^2fm^4$ to the lowest $2^+$ state
in $^8\mbox{Be}$. With the $Q \cdot Q$ interaction the two values are
respectively $46.40$ and $49.16$ $e^2fm^4$. One big difference between
the two interactions is the ratio of large to small space values for
corresponding $B(E2)$ values. In $^8\mbox{Be}$ the ratio of the large
sum to the small sum is $1.98$ for the ($x,y$) interaction whereas it
is much larger $3.28$ for the $Q \cdot Q$ interaction. There is much more
core polarization with the $Q \cdot Q$ interaction than with the ($x,y$)
interaction.

There have been many discussions concerning the correlation of summed
orbital $M1$ strength and the $B(E2)$ from the $J=0^+$ ground state to
the first $2^+$ state. The latter is an indication of the nuclear
deformation. We have noted that the calculated values of $B(E2)$ are
about the same in $^8\mbox{Be}$ and $^{10}\mbox{Be}$. Thus we would
expect the orbital $M1$ strengths in the two nuclei to be about the
same.

There is a certain `vagueness' in what is meant by `strength'. It is
clear that the experiments thus far sample only low energy strengths
up to about $6$ $MeV$ in heavy deformed nuclei \cite{zr} \cite{rich}.
Also some of the theories involve summed strength per se and others
involve the energy weighted strength. Rather than enter into deep
philosophical discussions about what is meant by strength, we will
give a variety of ratios of strength
$\frac{^{10}\mbox{Be}}{^8\mbox{Be}}$ in Table $V$. We see that all the
ratios, be they non-energy weighted or energy weighted, be they in
small spaces or in large spaces, are substantially less than one. In
forming the ratios, we added for the numerator ($^{10}\mbox{Be}$) the
$J=0^+$ $T=1$ to $J=1^+$ $T=1$ and $J=0^+$ $T=1$ to $J=1^+$ $T=2$ strengths.

\section{A comparison of the $J=1^+$ $\rightarrow$ $0_1^+$ and $J=1^+$
$\rightarrow$ $2_1^+$ Magnetic Dipole Transitions}

\hspace{.25 in}
Let us assume that the $0_1^+$ and $2_1^+$ states are members of a
$K=0$ rotational band and that the $1^+$ states have $K=1$. We can
then use the rotational formula of Bohr and Mottelson (Eq. 4-92) in
their book \cite {bm} ($K_1=0$, $K_2=1$) (We use the notation
$\left[\begin{array}{lll}J_1 & J_2 & J\\ M_1 & M_2 &
M\end{array}\right]$ for a Clebsch-Gordan coefficient):

\[\langle K_2I_2 || M(\lambda) || K_1=0 I_1 \rangle
=\sqrt{2(2I_1+1)}\left[\begin{array}{lll}I_1 & \lambda & I_2\\ 0 & K_2 &
K_2\end{array}\right]
\langle K_2 | M(\lambda,\nu=K_2)|K_1=0\rangle\]

\noindent From this we can easily deduce

\[r=\frac{B(M1)_{J=1^+, K=1 \rightarrow J=2^+, K=0}}{B(M1)_{J=1^+, K=1
\rightarrow J=0^+, K=0}} =\frac{1}{2}\]

\noindent Note, however, that the experimental ratio for $^8Be$ from
the $J=1^+$ $T=1$ state at 17.64 $MeV$ (see section 1) is
$\frac{0.12}{0.15}=0.8$. Bohr and Mottelson later discuss corrections
to the above simple formula.

We can obtain a value for the above ratio by forming an intrinsic
state and projecting out states of angular momentum $J=0$ and $J=2$.
We assume an extreme prolate shape for $^8Be$ and put the four valence
nucleons ($N\uparrow$, $N\downarrow$, $P\uparrow$, $P\downarrow$) in the
lowest Nilsson orbit with $\Lambda=0$. The asymptotic wave function is
$NrY_{10}$.

\noindent The $J=0$ wave function is

\[N'\sum_L \left[\begin{array}{lll}1 & 1 & L\\0 & 0 &
0\end{array}\right]  \left[\begin{array}{lll}1 & 1 & L\\0 & 0 &
0\end{array}\right]  \left[\begin{array}{lll}L & L & 0\\0 & 0 &
0\end{array}\right]  [L~L]^0 = 0.74536 [0~0]^0~+~0.66666 [2~2]^0\]

\noindent where the notation $[L_\pi~L_\nu]^J$ is used.

\noindent The $J=2$ wave function is

\[N''\left[\begin{array}{lll}1 & 1 & 0\\0 & 0 &
0\end{array}\right]  \left[\begin{array}{lll}1 & 1 & 2\\0 & 0 &
0\end{array}\right]  [0~2]^2~+~ \left[\begin{array}{lll}1 & 1 &
2\\0 & 0 &
0\end{array}\right]  \left[\begin{array}{lll}1 & 1 & 0\\0 & 0 &
0\end{array}\right] [2~0]^2\]

\[~+~  \left[\begin{array}{lll}1 & 1& 2\\0 & 0 &
0\end{array}\right]  \left[\begin{array}{lll}1 & 1 & 2\\0 & 0 &
0\end{array}\right] \left[\begin{array}{lll}2 & 2 & 2\\0 & 0 &
0\end{array}\right] [2~2]^2\]

\noindent i.e.

\[\psi^{J''=2}=0.62361[0~2]^2~+~0.62361[2~0]^2+0.47141[2~2]^2\]

We don't actually have to specify the $J=1^+$ state in detail to get
the ratio. We note that only the component $[2~2]^1$ of the $J=1$ wave
function can contribute to the $M1$ transition. The probability of
this component factors out in the ratio. With some additional Racah
algebra, we can show that $r=\frac{7}{8}=0.875$. This should be
compared with the value $r=\frac{1}{2}$ of the simple rotational
formula and with the experimental value $r_{exp}=0.8$. We see that we
get better agreement by using this projection method.

To complete the story, using results described in the next section, we
are able to deduce that for $^8Be$ $B(M1)_{1^+\rightarrow
2_1^+}$=$\frac{7}{4\pi}~\mu_N^2$

\section{Supermultiplet Scheme with a $Q \cdot Q$ interaction}
\subsection{A. Supermultiplet Scheme in $^8Be$}

\hspace{.25 in}
The $Q \cdot Q$ interaction that we have been using fits in nicely with the
$L-S$ supermultiplet scheme of Wigner \cite {wig}. For the $p$ shell
the unitary group $U(3)$ is relevant since there are three states:
$L=1$, $M=$1, 0 and -1. A very useful reference for this section is
the book by Hammermesh \cite {ham}.

If the Hamiltonian were a Casimir operator of $U(3)$ all states of a
given special symmetry $[f]=[f_1,~f_2,~f_3]$ would be
degenerate. For the case of $1p$ shell a state with a given particle
symmetry $[f_1,~f_2,~f_3]$ is identical to a quantum oscillator
symmetry state \cite{Har,Elliot} $(\lambda, \mu)=(f_1-f_2,f_2-f_3)$.
The states $(\lambda, \mu,L)$ are eigenstates of our $Q \cdot Q$
interaction which is a linear combination of the Casimir operator of
$SU(3)$ and an $L\cdot L$ interaction. The latter gives rise to a terminating
rotational $L(L+1)$ spectrum for states of different $L$ but with the
same $[f]$. Amusingly, as has been pointed out by many, one gets
identical bands in all $p$ shell nuclei with this model provided the
coefficient of $L\cdot L$ is fixed.

Unlike in the $s,~d$ shell, nothing new is added by using the quantum
numbers ($\lambda,~ \mu$) instead of [$f_1,~f_2,~f_3$] for $1p$ shell
states. This is because the number of different $M$ states availbale
for particles (3) coincides with the number of possible directions for
oscillator quanta $(a^{\dagger}_x,~a^{\dagger}_y,~a^{\dagger}_z)$ and
a single creation operator correponds to each particle.

In more detail, the Casimir operator is $\tilde{C_2}=Q\cdot Q~-3\vec{L} \cdot
\vec{L}$. Hence,

\begin{eqnarray*}
\langle-\chi Q\cdot Q\rangle_{\lambda~\mu~L} & = &\bar{\chi}\left[-\langle
\tilde{C_2}\rangle_{\lambda~\mu}~+~3L(L+1)\right]\\
& = &
\bar{\chi}\left[-4(\lambda^2+\mu^2+\lambda\mu+3(\lambda+\mu)+3L(L+1)\right]
\end{eqnarray*}
where $\bar{\chi}=\chi \frac{5b^4}{32\pi}$ with $b$ the harmonic oscillator
length parameter ($b^2=\frac{\hbar}{m\omega}$).
The magnetic dipole modes in the $L~S~T$ representation are:

\begin{tabbing}
$L=1$ $S=0$ $T=0$ \hspace{1.2 in} \= $L=0$ $S=1$ $T=1$ (isovector spin mode)\\
$L=1$ $S=0$ $T=1$ (scissors mode) \> $L=1$ $S=1$ $T=0$\\
$L=0$ $S=1$ $T=0$                 \> $L=1$ $S=1$ $T=1$\\
\end{tabbing}

With the $Q\cdot Q$ interaction that we have chosen, transitions from
the $L=0~S=0$ ground state in $^8Be$ to all of these modes except one
will vanish. The only surviving mode is the $L=1~S=0~T=1$ scissors
mode.

Let us give briefly the energies and some properties of the states in
$^8Be$ ($\bar{\chi}=0.1865$):

(a)  [f]=[4,0]    ($\lambda,\mu$)=(4,0)      Ground State Band

\begin{tabbing}
\hspace{0.5in}$L$ \hspace{0.5in} \= $S$ \hspace{0.5in} \= $T$
\hspace{0.5in} \= $\frac{E}{\bar{\chi}}$ \hspace{0.5in} \= $E^*(MeV)$
\hspace{0.5in} \\
\hspace{0.5in} 0 \hspace{0.5in} \> 0 \hspace{0.5in} \> 0
\hspace{0.5in} \> $-112$ \hspace{0.5in} \> 0 \hspace{0.5in} \\
\hspace{0.5in} 2 \hspace{0.5in} \> 0 \hspace{0.5in} \> 0
\hspace{0.5in} \> $-94$ \hspace{0.5in} \> 3.36 \hspace{0.5in} \\
\hspace{0.5in} 4 \hspace{0.5in} \> 0 \hspace{0.5in} \> 0
\hspace{0.5in} \> $-52$ \hspace{0.5in} \> 11.19 \hspace{0.5in} \\
\end{tabbing}

(b)  [f]=[3,1]    ($\lambda,\mu$)=(2,1) -contains the scissors mode
($L=1$, $S=0$, $T=1$).

Note that the ($S,T$) combinations (0,1), (1,0) and (1,1) are allowed.

\begin{tabbing}
\hspace{0.5in}$L$ \hspace{0.5in} \= $\frac{E}{\bar{\chi}}$
\hspace{0.5in} \= $E^*(MeV)$ \hspace{0.5in} \\
\hspace{0.5in} 1 \hspace{0.5in}  \> $-58$ \hspace{0.5in} \> 10.07
\hspace{0.5in} \\
\hspace{0.5in} 2 \hspace{0.5in}  \> $-46$ \hspace{0.5in} \> 12.31
\hspace{0.5in} \\
\hspace{0.5in} 3 \hspace{0.5in}  \> $-28$ \hspace{0.5in} \> 15.67
\hspace{0.5in} \\
\end{tabbing}

(c)  [f]=[2,2]    ($\lambda,\mu$)=(0,2)
The ($S,T$) combinations (0,0), (0,2), (2,0) and (1,1) are allowed.

\begin{tabbing}
\hspace{0.5in}$L$ \hspace{0.5in} \= $\frac{E}{\bar{\chi}}$
\hspace{0.5in} \= $E^*(MeV)$ \hspace{0.5in} \\
\hspace{0.5in} 0 \hspace{0.5in}  \> $-40$ \hspace{0.5in} \> 13.43
\hspace{0.5in} \\
\hspace{0.5in} 2 \hspace{0.5in}  \> $-22$ \hspace{0.5in} \> 16.78
\hspace{0.5in} \\
\end{tabbing}

(d)  [f]=[2,1,1]    ($\lambda,\mu$)=(1,0)
The ($S,T$) combinations (0,1), (1,0), (1,1), (1,2) and (2,1) are
allowed.

\begin{tabbing}
\hspace{0.5in}$L$ \hspace{0.5in} \= $\frac{E}{\bar{\chi}}$
\hspace{0.5in} \= $E^*(MeV)$ \hspace{0.5in} \\
\hspace{0.5in} 1 \hspace{0.5in}  \> $-10$ \hspace{0.5in} \> 19.02
\hspace{0.5in} \\
\end{tabbing}

Note that this supermultiplet also has a state with the quantum
numbers of the scissors mode $L=1~S=0~T=1$.

Some further comments are in order. The scissors mode state in
$^{156}Gd$, as a single band state originally discovered in electron
scattering \cite{bo}, was found when finer resolution ($\gamma,\gamma'$)
experiments were performed to consist of many states \cite{berg,bohle}.
This was a
beautiful example of intermediate structure. The supermultiplet scheme
here affords a concrete example of the origin of the intermediate
structure. Our scissors mode state at an energy of $-58~\bar{\chi}$ is
degenerate with an $L=1~S=1~T=1$ state. If we introduce spin-dependent
interactions the two states will admix and the degeneracy will be
removed. We will get intermediate structure.

\subsection{B. Supermultiplet Scheme in $^{10}Be$}

\hspace{.25 in}
We now give the energies and some properties of the states in
$^{10}Be$ ($\bar{\chi}=0.1286$):
(a)  [f]=[4,2]    ($\lambda,\mu$)=(2,2)   (includes ground state).

Allowed states:

\begin{tabbing}
\hspace{0.5in}$L$ \hspace{0.5in} \= $S$ \hspace{0.5in} \= $T$
\hspace{0.5in} \= $\frac{E}{\bar{\chi}}$ \hspace{0.5in} \= $E^*(MeV)$
\hspace{0.5in} \\
\hspace{0.5in} 0 \hspace{0.5in} \> 0 \hspace{0.5in} \> 1
\hspace{0.5in} \> $-96$ \hspace{0.5in} \> 0 \hspace{0.5in} \\
\hspace{0.5in} $2_1$ \hspace{0.5in} \> 0 \hspace{0.5in} \> 1
\hspace{0.5in} \> $-78$ \hspace{0.5in} \> 2.32 \hspace{0.5in} \\
\hspace{0.5in} $2_2$ \hspace{0.5in} \> 0 \hspace{0.5in} \> 1
\hspace{0.5in} \> $-78$ \hspace{0.5in} \> 2.32 \hspace{0.5in} \\
\hspace{0.5in} 3 \hspace{0.5in} \> 0 \hspace{0.5in} \> 1
\hspace{0.5in} \> $-60$ \hspace{0.5in} \> 4.63 \hspace{0.5in} \\
\hspace{0.5in} 4 \hspace{0.5in} \> 0 \hspace{0.5in} \> 1
\hspace{0.5in} \> $-36$ \hspace{0.5in} \> 7.72 \hspace{0.5in} \\
\end{tabbing}

Note that the $2_1^+$ and $2_2^+$ states are degenerate in this
scheme. This arises from the fact that, in a rotational picture, the
$K$ values that are allowed are $\mu,~\mu-2,~...$. Thus we have a
degeneracy of $J=2_1^+$ $K=0$ and $J=2_2^+$ $K=2$. This degeneracy
does not correspond to the experimental situation -the $2_1^+$ and
$2_2^+$ states are at $3.368~MeV$ and $5.960~MeV$ respectively -well
separated.

As a practical matter this degeneracy gives problems for the shell
model code $OXBASH$ \cite{oxbash}. Shell model routines often give
wrong answers for transition rates when the states involved are
degenerate. To overcome this difficulty, we have introduced weak
additional terms in the Hamiltonian to remove the degeneracy e.g. we
use a weak one-body spin-orbit interaction $-\xi \vec{l}\cdot\vec{s}$
with $\xi=0.1~MeV$.

(b)  Two degenerate bands [f]=[4,1,1]  ($\lambda,\mu$)=(3,0),
[f]=[3,3]  ($\lambda,\mu$)=(0,3)

Allowed states:

\begin{tabbing}
\hspace{0.5in}$L$ \hspace{0.5in} \= $S$ \hspace{0.5in} \= $T$
\hspace{0.5in} \= $\frac{E}{\bar{\chi}}$ \hspace{0.5in} \= $E^*(MeV)$
\hspace{0.5in} \\
\hspace{0.5in} 1 \hspace{0.5in} \> 1 \hspace{0.5in} \> 1
\hspace{0.5in} \> $-66$ \hspace{0.5in} \> 3.853 \hspace{0.5in} \\
\hspace{0.5in} 3 \hspace{0.5in} \> 1 \hspace{0.5in} \> 1
\hspace{0.5in} \> $-36$ \hspace{0.5in} \> 7.716 \hspace{0.5in} \\
\end{tabbing}

We get the low-lying $1^+$ states (one from [4,1,1] and one from
[3,3]). Note however that we have $L=1,~S=1$, hence the states cannot
be excited by either the orbital operator or the spin operator.

(c)  Band which contains the scissors mode [f]=[3,2,1]
($\lambda,\mu$)=(1,1)

\begin{tabbing}
\hspace{1.0in}$L$ \hspace{0.5in} \= $S$ \hspace{0.5in} \= $T$
\hspace{0.5in} \= $\frac{E}{\bar{\chi}}$ \hspace{0.5in} \= $E^*(MeV)$
\hspace{0.5in} \\
scissors mode 1 \hspace{0.5in} \> 0 \hspace{0.5in} \> 1
\hspace{0.5in} \> $-30$ \hspace{0.5in} \> 8.49 \hspace{0.5in} \\
scissors mode 1 \hspace{0.5in} \> 0 \hspace{0.5in} \> 2
\hspace{0.5in} \> $''$ \hspace{0.5in} \> $''$ \hspace{0.5in} \\
\hspace{1.0in} 1 \hspace{0.5in} \> 1 \hspace{0.5in} \> 1
\hspace{0.5in} \> $''$ \hspace{0.5in} \> $''$ \hspace{0.5in} \\
\hspace{1.0in} 1 \hspace{0.5in} \> 0 \hspace{0.5in} \> 2
\hspace{0.5in} \> $''$ \hspace{0.5in} \> $''$ \hspace{0.5in} \\
\hspace{1.0in} 1 \hspace{0.5in} \> 2 \hspace{0.5in} \> 1
\hspace{0.5in} \> $''$ \hspace{0.5in} \> $''$ \hspace{0.5in} \\
\end{tabbing}

\noindent There are also several $L=2$ states with
$\frac{E}{\bar{\chi}}=-18$ and $E^*=10.03~MeV$.

Note that the $L=1,~S=0$ scissors modes finally make their appearance.
There are two branches -one with isospin $T=1$ and one with isospin
$T=2$. These two scissors mode states are {\underline {degenerate}} in
energy in the supermultiplet scheme $^{10}Be$.

(d)  [f]=[2,2,2] ($\lambda,~\nu$)=(0,0)

\begin{tabbing}
\hspace{0.5in}$L$ \hspace{0.5in} \= $\frac{E}{\bar{\chi}}$
\hspace{0.5in} \= $E^*(MeV)$ \hspace{0.5in} \\
\hspace{0.5in} 0 \hspace{0.5in}  \> $0$ \hspace{0.5in} \> 12.34
\hspace{0.5in} \\
\end{tabbing}

The (S T) values are (0 1), (1 0), (1 2), (2 1), (0 4) and (4 0).

\subsection{B(M1) Transitions in the $SU(3)$ Scheme}

\hspace{.25 in}
For ${ ^8}$Be and $^{10}Be$
the strength for orbital M1 transitions from
$J^{\pi}=0^+$ to scissors mode states
can be
obtained in the $SU(3)$ scheme by observing that in the process
the ground state intrinsic state is
transformed to the corresponding intrinsic state of $1^+$ state.
For example in ${ ^8}$Be, the orbital isovector M1 operator,
$\mu=\sum_i l_i~ \tau_z^i=(L^\pi - L^{\nu})$ ,transforms
the four nucleon  intrinsic state
$(4,0)$
into the intrinsic state of $1^+ 1$ state
that is (21).
The operator
$(L^{\pi}_0 - L^{\nu}_0)$ is the generator of the scissors mode.
The orbital part of ground state and $1^+ 1$ state
of ${ ^8}$Be in $SU(3)$ scheme can be projected
out from the corresponding maximum weight intrinsic states
$\vert[f](\lambda,\mu)\epsilon,\Lambda ,\rho>$ where
$\epsilon=2\lambda+\mu$ and $\Lambda = \rho =\frac {1}{2} \mu$,
by using the following projection\cite{Har},
\begin{equation}
\vert[f](\lambda,\mu)K,L,M> =\frac {(2L+1)}{a(\lambda \mu L K)}
\int d \Omega D^{L}_{M,K}(\Omega) R(\Omega)
\vert[f](\lambda,\mu)\epsilon,\Lambda,\rho>.
\label{proj}
\end{equation}
Eq.(\ref{proj}) is a general equation for projecting out
orbital part of the wave function for a given L from a
$(\lambda, \mu)$ state.
In particular for ${ ^8}$Be we have,
\begin{equation}
\vert[4](40)0,0,0>=\frac{1}{a(4000)} \int d \Omega
D^{0}_{00}(\Omega) R(\Omega)
\vert[4](40)8,0,0>
\end{equation}
and
\begin{equation}
\vert[31](21)1,1,0>=\frac{3}{a(2111)}\int d \Omega
D^{1}_{01}(\Omega) R(\Omega)
\vert[31](21)5,\frac {1}{2},\frac {1}{2}>.
\end{equation}
Knowing the [f] representation of the $(\lambda, \mu)$
states one can look up the corresponding
conjugate charge spin states to get definite JT states.

The orbital magnetic transition strength B(M1) between these
states is calculated by a method similar to that outlined in
Appendix A of ref.\cite{reta} and is found to be

\begin{equation}
\begin{array}{c}
B(M1)=  \frac {9}{16 \pi} \vert <[31](21)110,S=0,T=1 \vert
(L^{\pi}_0 - L^{\nu}_0)
\vert[4](40)000, S=0, T=0>\vert^2\\
 = \frac {2}{\pi}\mu_n^2 =0.637~\mu_n^2
\end{array}
\end{equation}

In an analogous fashion one can calculate the
scissors mode  M1 transition strengths for the nucleus
$^{10}Be$. We find the following  results for SU(3) limit
transition strengths in $^{10}Be$:

\[B(M1)(0^+1 \rightarrow 1^+1)=\frac {9}{32 \pi}\mu_n^2
=0.0895\mu_n^2  \]
\[B(M1)(0^+1 \rightarrow 1^+2)=\frac {15}{32 \pi}\mu_n^2
=0.1492\mu_n^2 . \]
\subsection{Realistic Spin-Orbit Interaction and Restoration
of $SU(3)$ Symmetry}

\hspace{.25 in}
As pointed out before, an
important role of spin dependent part of interaction is
to remove the
degeneracies present in the $SU(3)$ limit by mixing up
the same final angular momentum states arising due to a given
intrinsic state as well as from different  intrinsic
states. In a realistic interaction, the relative strengths
of spin dependent and spin independent part of interaction
determine whether the wavefunctions are close
to SU(3) scheme or a (j-j) coupling scheme is a better
description of the system.

To understand further the part played by spin independent part
of the full realistic
interaction in the restoration of $SU(3)$ symmetry, we consider
a small space calculation with the full spin-orbit part of the
$(x,y)$ interaction plus a variable $Q.Q$ interaction. Figure.(7)
is a plot of isovector orbital, spin and total
strength for M1 transitions from
$J=0_1^+ T=0 \rightarrow J=1^+ T=1$ states versus $t$, the parameter
multiplying the full Q.Q interaction matrix elements for
${ ^8}$Be. For the spin part we use the operator 9.412$\sum \sigma
t_z$ i.e. we {\underline{include}} the large isovector factor.
In ${ ^8}$Be, with increasing $t $ the orbital isovector strength
is seen to approach the SU(3) limit value of $0.637$$\mu_n^2$.
The contribution of isovector spin transition, on the other
hand to total B(M1) decreases as $t$ becomes large. This is because
the $SU(4)$ limit is being approached and in this limit the spin
contribution vanishes.

In Fig.(7) with only spin-orbit part of realistic interaction in play,
the calculated
M1 transition strength for $0^+ 0$ to $1^+ 1$ transitions has a large
spin flip contribution and is found tobe as large as 9.7 $\mu_n^2$.
In a small space calculation with full realistic
interaction$(x,y)$(Table I), a total B(M1)
value of 1.0547$\mu_n^2$ is obtained with an isovector orbital
transition strength of 0.67$\mu_n^2$ and an isovector spin contribution
of 0.38$\mu_n^2$. Of course to get the physical $B(M1)$ we add the
spin and orbital {\underline{amplitudes}} and square. The spin $B(M1)$
is a factor of 25 lower here than the $t=0$ value in Fig 7. It still
has some effect because of the factor 9.412.
We may note that the orbital transition strength
arising due to full realistic interaction is very close to the
$SU(3)$ limit indicating that the realistic interaction
favors a restoration of $SU(3)$ symmetry. The large
space realistic interaction calculation inspite of the
correlations induced by shell mixings results in a B(M1)
value 1.2866$\mu_n^2$ and isovector orbital transition strength
of 0.728$\mu_n^2$
indicating that the wavefunctions are still
very close to $SU(3)$ wave functions.

In $ ^{10}$Be the situation is more interesting due to
splitting of scissors mode strength into two degenerate
states in SU(3) limit. Figures (8) and (9)
show  the orbital and spin part respectively of
M1 transition strength for transitions
from ground state to $J=1^+ T=1$ states,
$ J=1^+ T=2$ states and all $J=1^+$ states.
For ground state to $J=1^+ T=1$ transitions the orbital
B(M1) is seen to dip to a minimum for $t=0.3$
before it starts increasing so as to approach its
SU(3) limit. An opposite trend is observed in the
corresponding spin strength that shows some increase,
reaches a maximum and then tends to the $SU(4)$ limit of zero.
The characteristic behaviour at $t=0.3$ is possibly a manifestation of
a shape change at small deformation before the system
stabilizes by acquiring a permanent deformation.
The M1 transition sums for
ground state to $J=1^+ T=2$ states,
show a behaviour similar to that observed for
ground state to $J=1^+ T=1$ transitions in ${ ^8}$Be.

\section{Magnetic Dipole Transitions To Individual States}

\hspace{.25 in}
We here present several tables of magnetic dipole transitions from the
$J=0^+$ ground states of $^8Be$ and $^{10}Be$ to individual $J=1^+$
states. We use both the ($x,y$) interaction with $x=1,~y=1$ and the $Q
\cdot Q$ interaction.

Concerning the latter, we learned in the previous section that there
are many degenerate states in the $0\hbar\omega$ calculation when a $Q
\cdot Q$ interaction is used. Unfortunately, most shell model
routines, including the one used here, give erroneous results for
transition rates when there are degeneracies. In all our small space
($0\hbar\omega$) calculations using $Q \cdot Q$ we have added a small
spin-orbit interaction $-\xi \vec{l} \cdot \vec{s}$ with
$\xi=0.1~MeV$. This works but it introduces an artificial complexity
in our tables. However, it is easy to see by eye what states would be
degenerate if the spin-orbit interaction is removed. Alternatively,
one can use the analytic expressions for the energies in the previous
section.

The columns in Tables $VI$ through $XIII$ are defined as in Table $I$:

\begin{tabbing}
\hspace{2.5in}\= $g_{l\pi}$ \hspace{0.3in}\= $g_{l\nu}$\hspace{0.3in}
\= $g_{s\pi}$\hspace{0.3in} \= $g_{s\nu}$\\
(a) Isovector Orbital \> 0.5 \> -0.5 \> 0     \> 0    \\
(b) Isovector Spin    \> 0   \> 0    \> 0.5   \> -0.5 \\
(c) Physical          \> 1   \> 0    \> 5.586 \> -3.826\\
\end{tabbing}

\subsection{Calculated Magnetic Dipole Transitions in $^8Be$}

\hspace{0.25 in}
In Tables $VI$ and $VII$ we present the details of the $0\hbar\omega$
calculated $B(M1)$ values from the $J=0^+,~T=0$ ground state of $^8Be$
to the $J=1^+,~T=1$ excited states.

For the realistic ($x,y$) interaction, the isovector orbital strength
is concentrated in three states at 13.7, 16.6 and 18.0 $MeV$. The sum
of the orbital strengths is $\sum B(M1)\uparrow=0.67~\mu_{N}^2$. This is in
fair agreement with the experimental value
$B(M1)\uparrow=0.81~\mu_{N}^2$ (which is actually deduced from the
downward $\gamma$ decay of the 17.64 $MeV$ $J=1^+,~T=1$ state to the
ground state). However, in the experiment all the strength is
concentrated in {\underline {one}} state whereas in our calculation we
have considerable fragmentation. On the other hand, if we look at the
physical transitions, there is much more concentration in one state at
13.73 $MeV$ with $B(M1)\uparrow=0.72~\mu_{N}^2$. We will discuss this
more soon.

With the $Q \cdot Q$ interaction, all the orbital strength is
concentrated in the (2-fold degenerate) state at $10.1~MeV$ with a
summed strength $B(M1)\uparrow=0.64~\mu_{N}^2$, very similar to that
of the ($x,y$) interaction. The energy is too low compared with
experiment, but we must remember that we did not renormalize the
strength $\chi$ to allow for $\Delta N=2$ admixtures. Note that the
isovector spin transitions are zero with the $Q \cdot Q$ interaction.
This is because we are at the $SU(4)$ limit since $Q \cdot Q$ is a
spin-independent interaction.

Note that with the $Q \cdot Q$ interaction the summed orbital strength
is $\frac{2}{\pi}~\mu_{N}^2$, confirming the expressions that were
derived in the previous section. Coming back to the ($x,y$)
interaction, we see that here also the isovector spin transitions are
very weak. But note that for the 13.73 $MeV$ state whereas the orbital
value of $B(M1)\uparrow$ is 0.2569 $\mu_{N}^2$ and the spin value is
0.0013 $\mu_{N}^2$, the physical value is 0.7155 $\mu_{N}^2$. The
reason is that the spin and orbit amplitudes add coherently and that
the spin amplitude is multiplied by the factor $9.412$.

For other states there is destructive interference between spin and
orbit. For example, for the 16.64 $MeV$ state, the value of
$B(M1)\uparrow$ is 0.234 $\mu_{N}^2$ for the orbital case but only
0.065 $\mu_{N}^2$ for the physical case.

\subsection{Calculated Magnetic Dipole Transitions in $^{10}
Be$}

\hspace{0.25 in}
In Tables $VIII$ and $IX$ we present the details of the $0\hbar\omega$
calculated $B(M1)$ values from the $J=0^+~T=1$ ground state of
$^{10}Be$ to the $J=1^+~T=1$ and to $J=1^+~T=2$ excited states. We
caution the reader that whereas for $^8Be$ we presented the results in
units of $\mu_{N}^2$ (Tables $VI$ and $VII$), for $^{10}Be$ we use
$10^-2$ $\mu_{N}^2$ as the unit. The reason for this is that the
orbital transitions to {\underline {individual}} states in $^{10}Be$
are considerably smaller than those in $^8Be$.

Let us look at the $Q \cdot Q$ interaction (Table $IX$) first. There
are several outstanding features which are explained in the previous
section on $L~S$ coupling and supermultiplet symmetry.

The first two $J=1^+~T=1$ states are degenerate at $E^*=3.86$ $MeV$.
They carry no spin or orbital strength from the from the ground state.
The [f] symmetries are [4 1 1] and [3 3]. They have additional quantum
numbers $L=1~S=1~T=1$. Since the isovector orbital operator
($\vec{L_{\pi}}$-$\vec{L_{\nu}}$) cannot change both $L$ and $S$ from
zero to one, the orbital $B(M1)$ vanishes. A similar argument holds
for the isovector spin operator. These lowest two states are therefore
not scissors mode states.

Then we have a four fold set of degenerate states with [3 2 1]
symmetry at about 8.5 $MeV$ excitation which does include the
$L=1~S=0~T=1$ scissors mode. We note that for $^{10}Be$, the $T=1$
scissors mode is {\underline {degenerate}} with the $T=2$ scissors mode
also at 8.5 $MeV$ excitation. This again is a prediction of the
supermultiplet theory.

The summed isovector orbital strength is $\frac{9}{32\pi}$ $\mu_{N}^2$
from $J=0^+~T=1$ to the $J=1^+~T=1$ states, and it is
$\frac{15}{32\pi}$ $\mu_{N}^2$ to the $J=1^+~T=2$ states. We have in
effect a ($2T+1$) rule:

\[\frac{(2T+1)_{T=2}}{(2T+1)_{T=1}}=\frac{5}{3}\]

This coincides with the ratio of $T=2$ to $T=1$ strength.

Recalling that the $^8Be$ strength was $\frac{2}{\pi}$ $\mu_{N}^2$, we
see that the ratio of total strength $\frac{^{10}Be}{^8Be}$ is
$\frac{3}{8}$.

We now come back to Table $VIII$ which shows the same calculational
results with the `realistic' ($x,y$) interaction. There are several
similarities but also some differences with the $Q \cdot Q$ results.
Just as with the $Q \cdot Q$ interaction, the orbital transitions to
the lowest two $J=1^+$ $T=1$ states at 6.14 and 7.68 $MeV$ are very
weak 0.16 and 0.17 $\times~(10^{-2}\mu_{N}^2)$ respectively. However,
the spin transitions, which with $Q \cdot Q$ were also zero, are now
sufficiently strong so as to have a visible effect. For example, the
physical $B(M1)\uparrow$ to the 7.68 $MeV$ state is calculated to be
1.85 $\mu_{N}^2$. This is certainly measurable.

As with the $Q \cdot Q$ interaction, the scissors mode states with the
($x,y$) interaction are at a much higher energy than the lowest two
$1^+$ states ~ 19 $MeV$. Also, the $J=1^+$ $T=1$ and $T=2$ excitations are
roughly in the same energy range -the $Q \cdot Q$ interaction has them
degenerate. The ratio of $T=2$ to $T=1$ orbital strength is about the
same for the ($x,y$) interaction as for $Q \cdot Q$ 1.44 vs.
$\frac{5}{3}$.

One major difference is that the energy scale is larger for the
($x,y$) interaction than for $Q \cdot Q$. The lowest and higher
energies in Table $VIII$ are 6.14 $MeV$ and 30.96 $MeV$ whereas in
Table $IX$ they are 3.85 $MeV$ and 12.35 $MeV$.

In Tables X and XI we present results of large space
calculations for $^8Be$ to be compared with the corresponding small
space Tables VI and VII. Likewise in Tables XII and XIII we present
the large space results for $^{10}Be$ to be compared with tables VIII
and IX. We do not show all the states here, only
the low energy sector. The excitation energies are in general larger
in the large space calculations. The major changes occur when one has
nearly degenerate levels sharing some strength. For example, the
lowest two $1^+$ states in the large space calculations, which are at
7.38 $MeV$ and 9.62 $MeV$ have almost equal $M1$ strengths 0.64
$\mu_{N}^2$ and 0.79 $\mu_{N}^2$. In the small space, the lowest state
has only 0.011 $\mu_{N}^2$ and the second one 1.8 $\mu_{N}^2$.

A sensible attitude is to assume that neither calculation is accurate
enough to give the detailed distribution of strength between the two
states -only the summed strength for the two states should be compared
with experiment.

With the $Q \cdot Q$ interaction in the large space, the degeneracy
encountered in the small space calculation is removed. In part, this
is due to the fact that we do not include the single-particle terms
$\sum_{i=j}Q(i) \cdot Q(j)$. These will induce a single-particle
splitting between $1s$ and $0d$ in the $N=2$ shell. But since
degeneracies give us trouble in our shell model diagonalizations, we
are happy to leave the calculation as is.

With $Q \cdot Q$ the scissors mode strength in $^8Be$ gets pushed up
from the small space value of 10.1 $MeV$ to the large space value of
15.5 $MeV$. For $^{10}Be$ the corresponding numbers are 8.5 $MeV$ and
11.3 $MeV$ for the $J=1^+,~T=1$ states and essentially the same for
$J=1^+,~T=2$ states. That is, the near degeneracy of the $T=1$ and
$T=2$ scissors modes in $^{10}Be$ is maintained in the large space
calculation.

Lastly, we reiterate the fact that the shell model calculations here
yield not only colective magnetic states but also show intermediate
structure. That this structure is a natural occurence is shown by the
supermultiplet model, where for a given $[f]_{L=1}$ there are several
$S$ and $T$ values possible. It is of course very difficult to get the
details of the intermediate structure to come out right, but it is
good to be able to explain the origin of this structure.

\section{Additional Comments and Closing Remarks}

\hspace{.25 in}
We can gain further insight into the nature of $^8Be$ and $^{10}Be$ by
evaluating the quadrupole moments of the $J=2^+$ states. A small space
calculation gives the following values:

\begin{tabbing}
\hspace{1.8in} \= $Q \cdot Q~-0.1\vec{l}\cdot\vec{s}$ \hspace{.5in}\= ($x,y$)
interaction\\
$^8Be~~J=2^+,~T=0$ \> $Q=-8.02~e~fm^2$ \> $Q=-7.86~e~fm^2$\\
$^{10}Be~~J=2_1^+,~T=1$ \> $Q=-2.52~e~fm^2$ \>$Q=-7.68~e~fm^2$ \\
$^{10}Be~~J=2_2^+,~T=1$\> $Q=+2.06~e~fm^2$ \> $Q=+6.91~e~fm^2$\\
\end{tabbing}

In the rotational model the quadrupole moment of the $2^+$ of a $K=0$
band is $-\frac{2}{7}Q_0$ where $Q_0$ is the intrinsic quadrupole
moment. Thus a negative $Q$ corresponds to a prolate shape and a
positive $Q$ to an oblate shape.
{}From the above, $^8Be$ acts as a normal deformed nucleus of the
prolate shape.

It has been pointed out by Harvey that in
the $SU(3)$ scheme, whenever $\mu$ is less than or equal to $\lambda$
the nucleus becomes oblate \cite{Har}. For the ground state band in
$^8Be$ $\lambda$ is bigger than $\mu$ but for $^{10}Be$ $\lambda$ and
$\mu$ are equal. The situation with $^{10}Be$ is somewhat confusing.
With the $Q \cdot Q$ interaction, which one might think would favor
deformation, the quadrupole moment of the $2_1^+$ state drops to -2.52
$e~fm^2$. Recall that for a perfect vibrator, the value of $Q$ is
zero, so it would appear that $^{10}Be$ is headed in that direction.
However with the realistic interaction, which contains a large
spin-orbit interaction that one might think would oppose deformation,
the quadrupole moment of $^{10}Be$ becomes more negative -almost the
same as that of $^8Be$. Note also that the calculated values of $Q$
for the $2_1^+$ and $2_2^+$ states are nearly equal but opposite to
both interactions.

We have learned many interesting things by considering scissors modes
in light nuclei. First of all there is evidence for their existence.
This evidence comes strangely from a nucleus whose ground state is
unstable -$^8Be$. We learn of the existence from the inverse process
i.e. $\gamma$ decay from the $J=1^+,~T=1$ state at $17.64~MeV$ to the
ground state \cite{ajz}. The decay to the $2_1^+$ state, presumably a
member of a $K=0$ rotational band, is also observed and this suggests
that theoretical studies (and experimental ones as well whenever
possible) should be made not only between between $J=1^+~K=1$ and
$J=0^+~K=0$ states but also between $J=1^+~K=1$ and $J=2^+~K=0$
states. This will make the picture of scissors modes more complete. In
this work we considered but one example and showed that the ratio of
$J=1^+$ decay rate to $J=2^+$ vs $J=0^+$ deviates from the simple
rotational formula result of 0.5. Further studies along these lines
are being planned.

To make the picture even more complete, one can also study the decay
of $J=1^+~K=1$ to $J=2^+~K=2$. We were almost forced into such a study
by the fact that in the $SU(3)$ scheme there is a two-fold degeneracy
of the lowest $J=2^+$ states in $^{10}Be$ \cite{ham,Elliot}.
Presumably, these two states are admixtures of $2^+~K=0$ and
$2^+~K=2$.

The shell model approach used here \cite{oxbash} enables us
to study fragmentation or intermediate structure. We find for
example that with an electromagnetic probe there are, besides the
strong scissors mode states, some almost {\underline{`invisible'}}
states. These have separately substantial orbital contributions and
substantial spin contributions to the magnetic dipole excitations but
the physical $B(M1)$ is very small because of the destructive
interference of the spin and orbital amplitudes. For example, as seen
in Table X, in $^8Be$ we calculate that the low lying orbital strength
is shared almost equally between two states (at 18.0 $MeV$ and 20.8
$MeV$) -the strengths being 0.22 $\mu_N^2$ and 0.30 $\mu_N^2$
respectively. However, the physical $B(M1)$'s are 0.62  $\mu_N^2$ and
0.12 $\mu_N^2$. Thus one can miss considerable orbital strength into
these `invisible' sttaes if one uses only an electromagnetic probe.

Another thing we learn is that although spin excitations are strongly
suppressed they cannot be ignored. In the $SU(4)$ limit, the spin
matrix elements vanish and there is a clear tendency in our
calculations in that direction. However, since the isovector spin
operator is multiplied by a factor of 9.412, the spin and orbital
contributions tend to be on the same footing.

In the example of the above paragraph in the decay of the (calculated)
18.0 $MeV$ state, the orbital $B(M1)$ is only 0.22 $\mu_N^2$ but the
physical one which induces the spin contribution is 0.62 $\mu_N^2$. In
$^{10}Be$ the first two $J=1^+$ states are calculated to be excited
mainly by the spin operator and the $B(M1)$'s should be substantial ~
0.5 $\mu_N^2$. Yet in the $U(3)-SU(4)$ limit, these lowest two states
[f]=[4,1,1] and [3,3] should not be excited at all either by the spin
or by the orbital operators.

We have found the Wigner supermultiplet scheme \cite{wig} combined
with the $SU(3)$ scheme \cite{Har,Elliot} a very useful guide to the
more complicated shell model calculations that we have performed.
There is the added simplicity in the $p$ shell that there is a
one-to-one correpondence between a given [f] symmetry and the
$(\lambda,\mu)$ symmetries. Many interesting properties about scissors
modes can be literally read off the pages of the book by Hammermesh
\cite{ham}. For example, there is the fact that the $T=1$ and $T=2$
scissors mode states in $^{10}Be$ are degenerate in energy. This is an
exact result with the $Q \cdot Q$ interaction in a $p$ shell
calculation. Results very close to this are obtained with a realistic
interaction, but we frankly didn't notice this until we made an
$SU(3)$ analysis. Also the non-obvious fact that the lowest two $1^+$
states in $^{10}Be$ are not scissors mode states is made clear by such
an analysis.

Also the fact that scissors mode states everywhere, including
$^{156}Gd$, have intermediate structure \cite{berg,bohle} is made
clear by the supermultiplet scheme. For a given $L=1$ state there are
often several $S,~T$ combinations which are degenerate. The removal of
the degeneracy and the mixing of these states e.g. by a spin-orbit
interaction leads to fragmentation and intermediate structure.

By extending the shell model calculations to `large space' i.e. by
including $2\hbar\omega$ excitations, we were able to calculate the
cumulative energy weighted strength distribution for isovector orbital
excitations. The results which are shown in several figures are
characterized by a low energy rise followed by a second plateau. The
shapes of the distributions were similar for the two contrasting
interactions used here -the `realistic' short range interaction and
the schematic quadrupole-quadrupole interaction. The results were
compared witht the simple Nilsson model \cite{dz1,no} which predicts
that the energy-weighted sum at high energy (beyond the first plateau)
should equal the low energy rise i.e. the ratio
$\frac{total}{low~energy}$ should be 2:1. The actual calculated ratios
witht the ($x,y$) interaction are 1.75 for $^8Be$ and 2.52 for
$^{10}Be$. The corresponding numbers for the $Q \cdot Q$ interaction
were 1.37 and 2.33. We see that the Nilsson model is not bad as a
first orientation but there are fluctuations. The fact that the above
ratios are larger for the ($x,y$) interaction than for the $Q \cdot Q$
interaction may support the idea of Hamamoto and Nazarewicz \cite{hz}
that the symmetry energy will cause the ratio to increase. Our
calculations however do not support their claim that the high energy
part of the energy weighted orbital strength should
{\underline{always be much larger}} than the low energy part
-certainly not for a `normal' rotational nucleus like $^8Be$. For
$^{10}Be$ the ratio is however somewhat larger than the Nilsson model
prediction.

Whether this is due to the atypical properties of $^{10}Be$ mentioned
in the text or is a harbinger of what will happen for most other
nuclei remains to be seen. From an experimental point of view, it
would be helpful to have more data on $^{10}Be$.
Not only have no $J=1^+$ states been identified but neither has the
$J=4_1^+$ state been seen. The location of this state might help us
decide whether $^{10}Be$ is rotational or vibrational.

At any rate, we should examine a larger range of nuclei and look into
more detail about the symmetry energy in order to be able to make more
definitive statements about the systematics of the cumulative energy
weighted distributions. We note that the Zheng-Zamick sum rule
\cite{zz} is able to handle the divergent behaviour between $^8Be$ and
$^{10}Be$. This sum rule involves the difference between isoscalar and
isovector summed $B(E2)$ strength, whereas corresponding expressions
by Heyde and de Coster \cite{ibm} based on the $I.B.A.$ model
\cite{iach,diep} as well as empirical analyses \cite{zr}
involve only $B(E2)$ to the lowest $2^+$ states. Even here more
sharpening up is in order.

Our initial reason for studying lighter nuclei is that they would give
us insight into the behaviour of heavier nuclei, and we could carry
out more complete calculations in the low $A$ region. But then we
found many results which made light nuclei studies fascinating for
their own sake. One rather broad lesson we have learned in the light
nucleus study is that there can be considerable change in going from
one nucleus to the next, and perhaps in heavier nuclei too much effort
has been made to make the nuclei fit into a smooth pattern. For
example, we find that whereas in $^8Be$ the lowest $1^+$ state is
dominantly a scissors mode state, in $^{10}Be$ the lowest $1^+$ states
are not scissors mode states at all -they can only be reached by the
spin operator. We should perhaps be looking for more variety of
behaviour in heavier nuclei. Lastly the supermultiplet scheme which we
found extremely useful was originally thought to be of interest only
in light nuclei where the spin-orbit interaction is small relative to
the residual interaction. However, it is now being realized that even
in heavy nuclei -especially for superdeformed states this scheme may
once again be very relevant. This would make our light nuclear studies
all the more important.

\vspace{.5in}

{\large {\bf Acknowledgement}}

This work was supported by the Department of Energy Grant No.
DE-FG05-86ER-40299. S.S. Sharma would like to thank the Department of
Physics at Rutgers University for its hospitality and to acknowledge
financial support from $CNP_q$, Brazil.
We thank E. Moya de Guerra for familiarizing us with
her work in collaboration with J. Retamosa, J.M. Udias and A. Poves.
We thank I. Hamamoto for her interest. We also thank N. Sharma for
useful comments about matrix diagonalization.

\pagebreak
\begin{small}
\samepage{
\nopagebreak{
\noindent
\begin{center}
{\bf Table I}.
Summed Magnetic Dipole Strengths (in $\mu_N^2$).
\begin{tabular}{|l|c|c|c|c|}
\multicolumn{5}{c}{}\\ \hline
\multicolumn{5}{|c|}{The ($x,y$) interaction ($x=1,~y=1$)}\\\hline
$Space$ & $Isovector~Orbital^{(d)}$ & $Isovector~Spin^{(e)}$ &
$Isoscalar~Orbital^{(f)}$ & $Physical^{(g)}$\\ \hline
\multicolumn{5}{|c|}{$^8\mbox{Be}$ $J=0_1^+$ $T=0$ $\rightarrow$ $J=1^+$
$T=1$}\\ \hline
$Small^{(a)}$ & 0.6701 & 0.00427 &  & 1.0547\\ \hline
$Large^{(b)}$ & 0.7283 & 0.00622 &  & 1.2866\\ \hline
$Low~Large^{(c)}$ & 0.5890 & 0.00322 &  & 0.8999\\ \hline
\multicolumn{5}{|c|}{$^{10}\mbox{Be}$ $J=0_1^+$ $T=1$ $\rightarrow$ $J=1^+$
$T=1$}\\ \hline
$Small$ & 0.1112 & 0.0234 & 0.0245 & 2.0930\\ \hline
$Large$ & 0.1963 & 0.0208 & 0.0270 & 1.9517\\ \hline
$Low~Large$ & 0.1002 & 0.0187 & 0.0200 & 1.6070\\ \hline
\multicolumn{5}{|c|}{$^{10}\mbox{Be}$ $J=0_1^+$ $T=1$ $\rightarrow$ $J=1^+$
$T=2$}\\ \hline
$Small$ & 0.1508 & 0.00105 &  & 0.0597\\ \hline
$Large$ & 0.1830 & 0.00222 &  & 0.2276\\ \hline
$Low~Large$ & 0.1339 & 0.000928 &  & 0.0754\\ \hline\hline
\multicolumn{5}{|c|}{The $Q \cdot Q$ interaction}\\\hline
\multicolumn{5}{|c|}{$^8\mbox{Be}$ $J=0_1^+$ $T=0$ $\rightarrow$ $J=1^+$
$T=1$}\\ \hline
$Small$ & 0.6364 & 0.0000 &   & 0.6364 \\ \hline
$Large$ & 0.7408 & 0.0005 &   & 0.7858 \\ \hline
$Low~Large$ & 0.6593 & 0.0002 &   & 0.6775 \\ \hline
\multicolumn{5}{|c|}{$^{10}\mbox{Be}$ $J=0_1^+$ $T=1$ $\rightarrow$ $J=1^+$
$T=1$}\\ \hline
$Small$ & 0.0895 & 0.0001 &   & 0.0986 \\ \hline
$Large$ & 0.1788 & 0.0004 &  & 0.2044\\ \hline
$Low~Large$ & 0.0922 & 0.0000  &  & 0.0881 \\ \hline
\multicolumn{5}{|c|}{$^{10}\mbox{Be}$ $J=0_1^+$ $T=1$ $\rightarrow$
$J=1^+$ $T=2$}\\ \hline
$Small$ & 0.1492 & 0.0000 &  & 0.1486\\ \hline
$Large$ & 0.1744 & 0.0002 &  & 0.1950  \\ \hline
$Low~Large$ & 0.1513  & 0.0000 &  & 0.1617 \\ \hline
\end{tabular}
\end{center}
}
}
\end{small}

\pagebreak
{\bf Table I Captions}

\noindent(a) Small Space $(0s)^4(0p)^6$

\noindent(b) Large Space $(0s)^4(0p)^6$ $+$ all $2\hbar\omega$ excitations

\noindent(c) Low energy part of Large Space (up to the first plateau)

\noindent(d) $g_{l\pi}=0.5$ $g_{l\nu}=-0.5$ $g_{s\pi}=0$ $g_{s\nu}=0$

\noindent(e) $g_{l\pi}=0$ $g_{l\nu}=0$ $g_{s\pi}=0.5$ $g_{s\nu}=-0.5$

\noindent(f) $g_{l\pi}=0.5$ $g_{l\nu}=0.5$ $g_{s\pi}=0$ $g_{s\nu}=0$.
The value for isoscalar spin is the same as for isoscalar orbital in
the case of $^{10}Be$ $J=0^+,~T=1$ $\rightarrow$ $J=1^+,~T=1$. For
$\Delta T=1$ the isoscalar case gives zero.

\noindent(g) $g_{l\pi}=1$ $g_{l\nu}=0$ $g_{s\pi}=5.586$
$g_{s\nu}=-3.826$
\pagebreak

\begin{small}
\samepage{
\nopagebreak{
\noindent
\begin{center}
{\bf Table II}.
Summed Energy Weighted Magnetic Dipole Strengths (in $\mu_N^2 MeV$)
\begin{tabular}{|l|c|c|c|c|}
\multicolumn{5}{c}{}\\ \hline
\multicolumn{5}{|c|}{The ($x,y$) interaction ($x=1,~y=1$)}\\\hline
$Space$ & $Isovector~Orbital$ & $Isovector~Spin$ &
$Isoscalar~Orbital$ & $Physical$\\ \hline
\multicolumn{5}{|c|}{$^8\mbox{Be}$ $J=0_1^+$ $T=0$ $\rightarrow$ $J=1^+$
$T=1$}\\ \hline
$Small$ & 10.689 & 0.08359 &  & 16.878\\ \hline
$Large$ & 20.744 & 0.2789 &  & 44.092\\ \hline
$Low~Large$ & 11.854 & 0.0778 &  & 18.349\\ \hline\hline
\multicolumn{5}{|c|}{$^{10}\mbox{Be}$ $J=0_1^+$ $T=1$ $\rightarrow$ $J=1^+$
$T=1$}\\ \hline
$Small$ & 2.0750 & 0.1954 & 0.2328 & 18.852\\ \hline
$Large$ & 7.9511 & 0.3119 & 0.6929 & 35.909\\ \hline
$Low~Large$ & 2.4680 & 0.1814 & 0.2369 & 18.429\\ \hline\hline
\multicolumn{5}{|c|}{$^{10}\mbox{Be}$ $J=0_1^+$ $T=1$ $\rightarrow$ $J=1^+$
$T=2$}\\ \hline
$Small$ & 2.984 & 0.0229 &  & 1.4783\\ \hline
$Large$ & 6.720 & 0.1167 &  & 12.83\\ \hline
$Low~Large$ & 3.3590 & 0.0267 &  & 2.2843\\ \hline\hline
\multicolumn{5}{|c|}{The $Q \cdot Q$ interaction}\\\hline
\multicolumn{5}{|c|}{$^8\mbox{Be}$ $J=0_1^+$ $T=0$ $\rightarrow$ $J=1^+$
$T=1$}\\ \hline
$Small$ & 6.411 & 0.0000 &  & 6.411\\ \hline
$Large$ & 14.040 & 0.0229 &  & 15.861\\ \hline
$Low~Large$ & 10.282 & 0.0037 &  & 10.619\\ \hline
\multicolumn{5}{|c|}{$^{10}\mbox{Be}$ $J=0_1^+$ $T=1$ $\rightarrow$ $J=1^+$
$T=1$}\\ \hline
$Small$ & 0.7597 & 0.0004 &  & 0.7942\\ \hline
$Large$ & 3.8108 & 0.0138 &  & 4.8627\\ \hline
$Low~Large$ & 1.041 & 0.0029 &  & 0.9904\\ \hline
\multicolumn{5}{|c|}{$^{10}\mbox{Be}$ $J=0_1^+$ $T=1$ $\rightarrow$
$J=1^+$$T=2$}\\ \hline
$Small$ & 1.2661 & 0.0000 &  & 1.2619\\ \hline
$Large$ & 2.6019 & 0.0068 &  & 3.1687\\ \hline
$Low~Large$ & 1.712 & 0.0005 &  & 1.8335 \\ \hline
\end{tabular}
\end{center}
}
}
\end{small}

\pagebreak
\begin{small}
\samepage{
\nopagebreak{
\noindent
\begin{center}
{\bf Table III}.
The $B(E2)$ values in $e^2 fm^4$ in $^8Be$ with (a) The ($x,y$)
interaction with $x=1$, $y=1$ and (b) the $Q \cdot Q$ interaction.
\begin{tabular}{|l|c|c|}
\multicolumn{3}{c}{}\\ \hline
$Interaction$ & $x=1$, $y=1$ & $\frac{1}{2}m\omega^2r^2-\chi Q \cdot
Q$\\ \hline
\multicolumn{3}{|c|}{$J=0_1^+$ $T=0$ $\rightarrow$ $J=2^+$ $T=0$}\\
\hline
$2_1^+$ $Small$ & 17.59 & 18.13\\
$2_1^+$ $Large$ & 25.97 & 49.16\\
$Sum$ $Small$ & 17.82 & 18.13\\
$Sum$ $Large$ & 35.21 & 59.38\\
$Sum$ $Low$ $Large$ & 33.28 & 49.22\\ \hline
\multicolumn{3}{|c|}{$J=0_1^+$ $T=1$ $\rightarrow$ $J=2^+$ $T=1$}\\
\hline
$Sum$ $Small$ & 2.31 & 2.59\\
$Sum$ $Large$ & 13.47 & 22.44\\
$Sum$ $Low$ $Large$ & 2.10 & 13.51\\ \hline
\end{tabular}

{\bf Table IV}.
The $B(E2)$ values in $e^2 fm^4$ in $^{10}Be$ (large space) with (a)
The ($x,y$) interaction with $x=1$, $y=1$ and (b) the $Q \cdot Q$
interaction.
\begin{tabular}{|l|c|c|}
\multicolumn{3}{c}{}\\ \hline
$Interaction$ & $x=1$, $y=1$ & $\frac{1}{2}m\omega^2r^2-\chi Q \cdot
Q$\\ \hline
\multicolumn{3}{|c|}{$J=0_1^+$ $T=1$ $\rightarrow$ $J=2^+$ $T=1$}\\
\hline
$2_1^+$ $Small$ & 22.98 & 17.31\\
$2_2^+$ $Small$ &  0.30 &  7.67\\
$Sum$  & 23.28 & 24.98\\
$2_1^+$ $Large$ & 19.68 & 5.17\\
$2_2^+$ $Large$ & 3.22 & 41.23\\
$Sum$  & 22.90 & 46.40\\  \hline
$Sum$ $Small$ & 24.21 & 25.50\\
$Sum$ $Large$ & 39.74 & 74.70\\
$Sum$ $Low$ $Large$ & 24.55 & 72.14\\ \hline
\multicolumn{3}{|c|}{$J=0_1^+$ $T=1$ $\rightarrow$ $J=2^+$ $T=2$}\\
\hline
$Sum$ $Small$ & 0.7858 & 0.8016\\
$Sum$ $Large$ & 5.851 & 12.19\\
$Sum$ $Low$ $Large$ & 0.5669 & 10.57\\ \hline
\end{tabular}
\end{center}
}
}
\end{small}

\pagebreak

\begin{small}
\samepage{
\nopagebreak{
\noindent
\begin{center}
{\bf Table V}.
Ratio of orbital $M1$ strength and orbital energy weighted $M1$ strength
$\frac{^{10}\mbox{Be}}{^8\mbox{Be}}$ for the ($x,y$) interaction with
$x=1$, $y=1$.
\begin{tabular}{|l|c|c|}
\multicolumn{3}{c}{}\\ \hline
 Space & Ratio of orbital & Ratio of energy weighted \\
       & strength & orbital strength \\ \hline
 $Small$ & 0.3910 & 0.4730\\
 $Large$ & 0.5208 & 0.7072\\
 $Low$ $Large$ & 0.3974 & 0.4920\\ \hline
\end{tabular}
\end{center}
}
}
\end{small}
\begin{small}
\samepage{
\nopagebreak{
\noindent
\begin{center}
{\bf Table VI}.
Calculated Magnetic Dipole Strength in $^8Be$ with the $(x,y)$
interaction $(x=1,~y=1)$ in small space (in units of $\mu_N^2$).
\begin{tabular}{|l|c|c|c|}
\multicolumn{4}{c}{}\\ \hline
$E_x(1^+)~MeV$ & $Isovector~Orbital$ & $Isovector~Spin$ &
$Physical$\\ \hline
\multicolumn{4}{|c|}{$J^{\pi}=0_1^+$ $T=1$ $\rightarrow$ $J^{\pi}=1^+$
$T=1$}\\ \hline
 13.73 & 0.2569 & 0.0013 & 0.7155\\
 16.64 & 0.2344 & 0.0006 & 0.0651\\
 18.05 & 0.1748 & 0.0002 & 0.0778\\
 20.73 & 0.0013 & 0.0012 & 0.1307\\
 26.80 & 0.0007 & 0.0005 & 0.0319\\
 28.46 & 0.0007 & 0.0004 & 0.0272\\
 29.66 & 0.0014 & 0.0000 & 0.0001\\
 34.76 & 0.0000 & 0.0001 & 0.0065\\\hline
  SUM  & 0.6701 & 0.0043 & 1.0547\\\hline
\end{tabular}
\end{center}
}
}
\end{small}

\pagebreak

\begin{small}
\samepage{
\nopagebreak{
\noindent
\begin{center}
{\bf Table VII}.
Calculated Magnetic Dipole Strength in $^8Be$ with the Hamiltonian
$H=\frac{p^2}{2m}~+~\frac{1}{2}m\omega^2r^2~-~\chi Q \cdot Q$ with
$\chi=0.5762$ in small space ($\mu_N^2$).
\begin{tabular}{|l|c|c|c|}
\multicolumn{4}{c}{}\\ \hline
$E_x(1^+)~MeV$ & $Isovector~Orbital$ & $Isovector~Spin$ &
$Physical$\\ \hline
\multicolumn{4}{|c|}{$J^{\pi}=0_1^+$ $T=1$ $\rightarrow$ $J^{\pi}=1^+$
$T=1$}\\ \hline
 10.06 & 0.4599 & 0.0000 & 0.4851\\
 10.11 & 0.1765 & 0.0000 & 0.1524\\
 12.33 & 0.0000 & 0.0000 & 0.0001\\
 13.43 & 0.0000 & 0.0000 & 0.0000\\
 16.84 & 0.0000 & 0.0000 & 0.0000\\
 18.97 & 0.0000 & 0.0000 & 0.0000\\
 19.05 & 0.0000 & 0.0000 & 0.0000\\
 19.11 & 0.0000 & 0.0000 & 0.0000\\\hline
  SUM  & 0.6364$^a$ & 0.0000 & 0.6376\\\hline
\end{tabular}
\end{center}
(a) The summed isovector orbital strength is $\frac{2}{\pi}~\mu_N^2$.
}
}
\end{small}

\pagebreak

\begin{small}
\samepage{
\nopagebreak{
\noindent
\begin{center}
{\bf Table VIII}.
Calculated Magnetic Dipole Strength in $^{10}Be$ with the $(x,y)$
interaction $(x=1,~y=1)$ in small space (in units of
$10^{-2}~\mu_N^2$).
\begin{tabular}{|l|c|c|c|}
\multicolumn{4}{c}{}\\ \hline
$E_x(1^+)~MeV$ & $Isovector~Orbital$ & $Isovector~Spin$ &
$Physical$\\ \hline
\multicolumn{4}{|c|}{A. $J=0_1^+$ $T=1$ $\rightarrow$ $J=1^+$
$T=1$}\\ \hline
 6.14 & 0.1641 & 0.0578 & 1.0740\\
 7.68 & 0.1699 & 2.1690 & 184.90\\
14.31 & 0.1000 & 0.0048 & 1.2020\\
16.48 & 0.0000 & 0.0105 & 1.1050\\
18.14 & 7.2840 & 0.0020 & 10.530\\
19.19 & 1.7540 & 0.0152 & 4.8420\\
22.62 & 1.1930 & 0.0064 & 0.4793\\
23.93 & 0.3904 & 0.0513 & 2.6420\\
25.72 & 0.0606 & 0.0231 & 2.5720\\ \hline
  SUM & 11.120 & 2.3409 & 209.30\\\hline
\multicolumn{4}{|c|}{B. $J=0_1^+$ $T=1$ $\rightarrow$ $J=1^+$
$T=2$}\\ \hline
17.91 & 6.4260 & 0.0331 & 0.6757\\
20.67 & 7.1290 & 0.0494 & 0.3339\\
23.50 & 1.5280 & 0.0039 & 3.3170\\
30.96 & 0.0001 & 0.0188 & 1.6430\\ \hline
  SUM & 15.083 & 0.1052 & 5.9696\\\hline
\end{tabular}
\end{center}
}
}
\end{small}
\pagebreak

\begin{small}
\samepage{
\nopagebreak{
\noindent
\begin{center}
{\bf Table IX}.
Calculated Magnetic Dipole Strength in $^{10}Be$ with the Hamiltonian
$H=\frac{p^2}{2m}~+~\frac{1}{2}m\omega^2r^2~-~\chi Q \cdot Q$ with
$\chi=0.3615$ in small space ($10^{-2}~\mu_N^2$).
\begin{tabular}{|l|c|c|c|}
\multicolumn{4}{c}{}\\ \hline
$E_x(1^+)~MeV$ & $Isovector~Orbital$ & $Isovector~Spin$ &
$Physical$\\ \hline
\multicolumn{4}{|c|}{A. $J=0_1^+$ $T=1$ $\rightarrow$ $J=1^+$
$T=1$}\\ \hline
 3.848 & 0.0006 & 0.0051 & 0.4173 \\
 3.870 & 0.0006 & 0.0046 & 0.5161\\
 8.444 & 3.0350 & 0.0001 & 2.7460\\
 8.486 & 0.0256 & 0.0000 & 0.0206\\
 8.491 & 3.3300 & 0.0000 & 3.3190\\
 8.533 & 2.5590 & 0.0001 & 2.8440\\
 9.988 & 0.0002 & 0.0000 & 0.0001\\
10.035 & 0.0003 & 0.0000 & 0.0003\\
10.082 & 0.0001 & 0.0000 & 0.0002\\\hline
  SUM  & 8.9555$^a$ & 0.0099 & 9.8636\\\hline
\multicolumn{4}{|c|}{B. $J=0_1^+$ $T=1$ $\rightarrow$ $J=1^+$
$T=2$}\\ \hline
 8.441 & 7.6680 & 0.0003 & 6.8270\\
 8.536 & 7.2500 & 0.0002 & 8.0320\\
10.034 & 0.0000 & 0.0000 & 0.0000\\
12.349 & 0.0000 & 0.0000 & 0.0000\\\hline
  SUM  & 14.918$^a$ & 0.0005 & 14.859\\\hline
\end{tabular}
\end{center}
(a) The summed isovector orbital strength to $T=1$ states is
$\frac{9}{32\pi}$ $\mu_N^2$, and to $T=2$ states it is
$\frac{15}{32\pi}$ $\mu_N^2$.
}
}
\end{small}
\pagebreak

\begin{small}
\samepage{
\nopagebreak{
\noindent
\begin{center}
{\bf Table X}.
Calculated Magnetic Dipole Strength in $^8Be$ with the $(x,y)$
interaction $(x=1,~y=1)$ in large space$^{(a)}$ (in units of
$\mu_N^2$).
\begin{tabular}{|l|c|c|c|}
\multicolumn{4}{c}{}\\ \hline
$E_x(1^+)~MeV$ & $Isovector~Orbital$ & $Isovector~Spin$ &
$Physical$\\ \hline
\multicolumn{4}{|c|}{$J=0_1^+$ $T=0$ $\rightarrow$ $J=1^+$
$T=1$}\\ \hline
 17.96 & 0.2208 & 0.0011 & 0.6160\\
 20.80 & 0.2970 & 0.0004 & 0.1213\\
 22.89 & 0.0670 & 0.0001 & 0.0213\\
 27.04 & 0.0000 & 0.0010 & 0.0935\\
 33.51 & 0.0007 & 0.0002 & 0.0081\\
 34.57 & 0.0000 & 0.0003 & 0.0250\\
 35.73 & 0.0007 & 0.0000 & 0.0000\\
 40.22 & 0.0001 & 0.0000 & 0.0036\\
 46.07 & 0.0003 & 0.0000 & 0.0110\\
 46.41 & 0.0000 & 0.0000 & 0.0000\\
 48.46 & 0.0060 & 0.0001 & 0.0001\\
 49.61 & 0.0019 & 0.0000 & 0.0023\\
 49.92 & 0.0012 & 0.0000 & 0.0004\\
 50.32 & 0.0004 & 0.0000 & 0.0036\\\hline
SUM & 0.5986 & 0.0033 & 0.9063\\\hline
\end{tabular}
\end{center}
\noindent(a) Large Space: all $0\hbar\omega$ configurations plus
$2\hbar\omega$ excitations.
}
}
\end{small}
\pagebreak

\begin{small}
\samepage{
\nopagebreak{
\noindent
\begin{center}
{\bf Table XI}.
Calculated Magnetic Dipole Strength in $^8Be$ with the Hamiltonian
$H=\frac{p^2}{2m}~+~\frac{1}{2}m\omega^2r^2~-~\chi Q \cdot Q$ with
$\chi=0.5762$ in large space ($\mu_N^2$).
\begin{tabular}{|l|c|c|c|}
\multicolumn{4}{c}{}\\ \hline
$E_x(1^+)~MeV$ & $Isovector~Orbital$ & $Isovector~Spin$ &
$Physical$\\ \hline
\multicolumn{4}{|c|}{$J=0^+$ $T=0$ $\rightarrow$ $J=1^+$ $T=1$}\\
\hline
15.49 & 0.6012 & 0.0000 & 0.5513 \\
15.70 & 0.0552 & 0.0001 & 0.1189\\
18.42 & 0.0003 & 0.0000 & 0.0005\\
22.69 & 0.0000 & 0.0000 & 0.0020\\
25.43 & 0.0000 & 0.0000 & 0.0004\\
27.82 & 0.0000 & 0.0000 & 0.0014\\
28.14 & 0.0000 & 0.0000 & 0.0002\\
28.24 & 0.0000 & 0.0000 & 0.0003\\
35.24 & 0.0009 & 0.0000 & 0.0013\\
37.88 & 0.0003 & 0.0000 & 0.0000\\
39.30 & 0.0001 & 0.0000 & 0.0001\\
39.51 & 0.0003 & 0.0000 & 0.0003\\
40.17 & 0.0010 & 0.0000 & 0.0010\\\hline
SUM & 0.6593 & 0.0002 & 0.6775\\\hline
\end{tabular}
\end{center}
}
}
\end{small}
\pagebreak

\begin{small}
\samepage{
\nopagebreak{
\noindent
\begin{center}
{\bf Table XII}.
Calculated Magnetic Dipole Strength in $^{10}Be$ with the $(x,y)$
interaction $(x=1,~y=1)$ in large space$^{(a)}$ (in units of
$10^{-2}~\mu_N^2$).
\begin{tabular}{|l|c|c|c|}
\multicolumn{4}{c}{}\\ \hline
$E_x(1^+)~MeV$ & $Isovector~Orbital$ & $Isovector~Spin$ &
$Physical$\\ \hline
\multicolumn{4}{|c|}{A. $J=0_1^+$ $T=1$ $\rightarrow$ $J=1^+$
$T=1$}\\ \hline
  7.38 & 0.006 & 0.909 & 63.78\\
  9.62 & 0.223 & 0.847 & 79.47\\
 18.29 & 0.015 & 0.000 & 0.026\\
 20.65 & 0.047 & 0.010 & 0.768\\
 22.73 & 6.087 & 0.004 & 10.40\\
 23.99 & 1.620 & 0.013 & 4.401\\
 27.15 & 1.411 & 0.004 & 0.788\\
 28.90 & 0.086 & 0.047 & 3.528\\
 30.86 & 0.017 & 0.031 & 2.831\\
 37.20 & 0.450 & 0.001 & 0.849\\
 39.96 & 0.002 & 0.003 & 0.150\\
 40.39 & 0.058 & 0.002 & 0.028\\\hline
  SUM  & 10.02 & 1.870 & 160.7\\\hline
\multicolumn{4}{|c|}{B. $J=0_1^+$ $T=1$ $\rightarrow$ $J=1^+$
$T=2$}\\ \hline
 23.27 & 4.733 & 0.015 & 1.058\\
 25.99 & 7.560 & 0.046 & 0.524\\
 29.11 & 1.004 & 0.013 & 4.311\\
 38.33 & 0.000 & 0.017 & 1.524\\
 49.42 & 0.017 & 0.000 & 0.098\\
 49.67 & 0.003 & 0.000 & 0.028\\
 50.84 & 0.075 & 0.000 & 0.000\\\hline
  SUM  & 13.39 & 0.093 & 7.540\\\hline
\end{tabular}
\end{center}
\noindent(a) Large Space: all $0\hbar\omega$ configurations plus
$2\hbar\omega$ excitations.
}
}
\end{small}
\pagebreak

\begin{small}
\samepage{
\nopagebreak{
\noindent
\begin{center}
{\bf Table XIII}.
Calculated Magnetic Dipole Strength in $^{10}Be$ with the Hamiltonian
$H=\frac{p^2}{2m}~+~\frac{1}{2}m\omega^2r^2~-~\chi Q \cdot Q$ with
$\chi=0.3615$ in large space ($10^{-2}~\mu_N^2$).
\begin{tabular}{|l|c|c|c|}
\multicolumn{4}{c}{}\\ \hline
$E_x(1^+)~MeV$ & $Isovector~Orbital$ & $Isovector~Spin$ &
$Physical$\\ \hline
\multicolumn{4}{|c|}{A. $J=0_1^+$ $T=1$ $\rightarrow$ $J=1^+$
$T=1$}\\ \hline
  3.60 & 0.002 & 0.000 & 0.019\\
  6.45 & 0.005 & 0.001 & 0.058\\
 11.19 & 2.548 & 0.000 & 2.393\\
 11.31 & 1.685 & 0.000 & 2.229\\
 11.32 & 4.563 & 0.001 & 3.382\\
 11.35 & 0.395 & 0.001 & 0.702\\
 12.92 & 0.019 & 0.000 & 0.019\\
 12.98 & 0.004 & 0.000 & 0.003\\
 13.02 & 0.002 & 0.000 & 0.003\\
 22.69 & 0.011 & 0.001 & 0.154\\
 23.12 & 0.019 & 0.000 & 0.001\\
 25.42 & 1.579 & 0.000 & 1.220\\
 25.94 & 0.043 & 0.003 & 0.497\\
 26.51 & 0.090 & 0.000 & 0.012\\
 26.65 & 0.004 & 0.000 & 0.005\\\hline
  SUM  & 10.97 & 0.007 & 10.70\\\hline
\multicolumn{4}{|c|}{B. $J=0_1^+$ $T=1$ $\rightarrow$ $J=1^+$
$T=2$}\\ \hline
 11.31 & 13.82 & 0.001 & 15.79\\
 11.34 & 1.301 & 0.003 & 0.348\\
 12.98 & 0.000 & 0.000 & 0.002\\
 16.76 & 0.000 & 0.000 & 0.007\\
 26.58 & 0.002 & 0.000 & 0.001\\
 26.67 & 0.002 & 0.000 & 0.010\\
 28.25 & 0.001 & 0.000 & 0.003\\\hline
  SUM  & 15.13 & 0.004 & 16.16\\\hline
\end{tabular}
\end{center}
}
}
\end{small}
\pagebreak

{\large{\bf {Figure Captions}}}

{\bf Figure (1):} The cumulative sum of the energy-weighted isovector
orbital $B(M1)$ strength for the $0_1^+,~0~\rightarrow~1^+,1$
transitions in $^8Be$ with the realistic interaction ($x=1,~y=1$).

{\bf Figure (2):} Same as Figure 1 but with the $Q \cdot Q$ interaction.

{\bf Figure (3):} The cumulative sum of the energy-weighted isovector
orbital $B(M1)$ strength for the $0_1^+,~1~\rightarrow~1^+,2$
transitions in $^{10}Be$ with the realistic interaction ($x=1,~y=1$).

{\bf Figure (4):} Same as Figure 3 but with the $Q \cdot Q$
interaction.

{\bf Figure (5):} The cumulative sum of the energy-weighted isovector
orbital $B(M1)$ strength for the $0_1^+,~1~\rightarrow~1^+,1$
transitions in $^{10}Be$ with the realistic interaction ($x=1,~y=1$).

{\bf Figure (6):} Same as Figure 5 but with the $Q \cdot Q$
interaction.

{\bf Figure (7):} $\sum B(M1)(J=0_1^+ T=0 \rightarrow
J=1^+ T=1$ states) versus $t$ for $^8Be$. The
solid line, dashed line and dot-dash line are
the total, spin and orbital parts of $\sum B(M1)$ .

{\bf Figure (8):} The orbital part of $\sum B(M1)$
versus $t$ for $ ^{10}Be$. The
solid line, dashed line and dot-dash line are
the total, $\sum B(M1)(J=0_1^+ T=1 \rightarrow
J=1^+ T=1$ states) and
$\sum B(M1)(J=0_1^+ T=1 \rightarrow J=1^+ T=2$ states)
respectively.

{\bf Figure (9):} Same as in Fig.(8) for spin part of $\sum B(M1)$.

\pagebreak
\begin{small}

\end{small}

\pagebreak

\end{document}